\RequirePackage{lineno}
\documentclass[twocolumn,showpacs,aps,superscriptaddress]{revtex4}

\usepackage{graphicx}
\usepackage{dcolumn}
\usepackage{amsmath}
\usepackage{supertabular}
\usepackage{multirow}
\usepackage{multicols}

\begin{document}


\title{ Energy and system-size dependence of two- and four-particle $v_2$ measurements in heavy-ion collisions at RHIC and their implications on flow fluctuations and nonflow}

\affiliation{Argonne National Laboratory, Argonne, Illinois 60439, USA}
\affiliation{Brookhaven National Laboratory, Upton, New York 11973, USA}
\affiliation{University of California, Berkeley, California 94720, USA}
\affiliation{University of California, Davis, California 95616, USA}
\affiliation{University of California, Los Angeles, California 90095, USA}
\affiliation{Universidade Estadual de Campinas, Sao Paulo, Brazil}
\affiliation{University of Illinois at Chicago, Chicago, Illinois 60607, USA}
\affiliation{Creighton University, Omaha, Nebraska 68178, USA}
\affiliation{Czech Technical University in Prague, FNSPE, Prague, 115 19, Czech Republic}
\affiliation{Nuclear Physics Institute AS CR, 250 68 \v{R}e\v{z}/Prague, Czech Republic}
\affiliation{University of Frankfurt, Frankfurt, Germany}
\affiliation{Institute of Physics, Bhubaneswar 751005, India}
\affiliation{Indian Institute of Technology, Mumbai, India}
\affiliation{Indiana University, Bloomington, Indiana 47408, USA}
\affiliation{Alikhanov Institute for Theoretical and Experimental Physics, Moscow, Russia}
\affiliation{University of Jammu, Jammu 180001, India}
\affiliation{Joint Institute for Nuclear Research, Dubna, 141 980, Russia}
\affiliation{Kent State University, Kent, Ohio 44242, USA}
\affiliation{University of Kentucky, Lexington, Kentucky, 40506-0055, USA}
\affiliation{Institute of Modern Physics, Lanzhou, China}
\affiliation{Lawrence Berkeley National Laboratory, Berkeley, California 94720, USA}
\affiliation{Massachusetts Institute of Technology, Cambridge, MA 02139-4307, USA}
\affiliation{Max-Planck-Institut f\"ur Physik, Munich, Germany}
\affiliation{Michigan State University, East Lansing, Michigan 48824, USA}
\affiliation{Moscow Engineering Physics Institute, Moscow Russia}
\affiliation{NIKHEF and Utrecht University, Amsterdam, The Netherlands}
\affiliation{Ohio State University, Columbus, Ohio 43210, USA}
\affiliation{Old Dominion University, Norfolk, VA, 23529, USA}
\affiliation{Panjab University, Chandigarh 160014, India}
\affiliation{Pennsylvania State University, University Park, Pennsylvania 16802, USA}
\affiliation{Institute of High Energy Physics, Protvino, Russia}
\affiliation{Purdue University, West Lafayette, Indiana 47907, USA}
\affiliation{Pusan National University, Pusan, Republic of Korea}
\affiliation{University of Rajasthan, Jaipur 302004, India}
\affiliation{Rice University, Houston, Texas 77251, USA}
\affiliation{Universidade de Sao Paulo, Sao Paulo, Brazil}
\affiliation{University of Science \& Technology of China, Hefei 230026, China}
\affiliation{Shandong University, Jinan, Shandong 250100, China}
\affiliation{Shanghai Institute of Applied Physics, Shanghai 201800, China}
\affiliation{SUBATECH, Nantes, France}
\affiliation{Texas A\&M University, College Station, Texas 77843, USA}
\affiliation{University of Texas, Austin, Texas 78712, USA}
\affiliation{University of Houston, Houston, TX, 77204, USA}
\affiliation{Tsinghua University, Beijing 100084, China}
\affiliation{United States Naval Academy, Annapolis, MD 21402, USA}
\affiliation{Valparaiso University, Valparaiso, Indiana 46383, USA}
\affiliation{Variable Energy Cyclotron Centre, Kolkata 700064, India}
\affiliation{Warsaw University of Technology, Warsaw, Poland}
\affiliation{University of Washington, Seattle, Washington 98195, USA}
\affiliation{Wayne State University, Detroit, Michigan 48201, USA}
\affiliation{Institute of Particle Physics, CCNU (HZNU), Wuhan 430079, China}
\affiliation{Yale University, New Haven, Connecticut 06520, USA}
\affiliation{University of Zagreb, Zagreb, HR-10002, Croatia}

\author{G.~Agakishiev}\affiliation{Joint Institute for Nuclear Research, Dubna, 141 980, Russia}
\author{M.~M.~Aggarwal}\affiliation{Panjab University, Chandigarh 160014, India}
\author{Z.~Ahammed}\affiliation{Variable Energy Cyclotron Centre, Kolkata 700064, India}
\author{A.~V.~Alakhverdyants}\affiliation{Joint Institute for Nuclear Research, Dubna, 141 980, Russia}
\author{I.~Alekseev~~}\affiliation{Alikhanov Institute for Theoretical and Experimental Physics, Moscow, Russia}
\author{J.~Alford}\affiliation{Kent State University, Kent, Ohio 44242, USA}
\author{B.~D.~Anderson}\affiliation{Kent State University, Kent, Ohio 44242, USA}
\author{C.~D.~Anson}\affiliation{Ohio State University, Columbus, Ohio 43210, USA}
\author{D.~Arkhipkin}\affiliation{Brookhaven National Laboratory, Upton, New York 11973, USA}
\author{G.~S.~Averichev}\affiliation{Joint Institute for Nuclear Research, Dubna, 141 980, Russia}
% ...

\collaboration{STAR Collaboration}\noaffiliation



\begin{abstract} %
  We present STAR measurements of azimuthal anisotropy by means of the two- and four-particle cumulants
  $v_2$ ($v_2\{2\}$ and $v_2\{4\}$) for Au+Au and Cu+Cu collisions
  at center of mass energies $\sqrt{s_{_{\mathrm{NN}}}} = 62.4$ and
  200~GeV. The difference between $v_2\{2\}^2$ and $v_2\{4\}^2$ is
  related to $v_{2}$ fluctuations ($\sigma_{v_2}$) and nonflow
  $(\delta_{2})$. We present an upper limit to
  $\sigma_{v_2}/v_{2}$. Following the assumption that eccentricity
  fluctuations $\sigma_{\varepsilon}$ dominate $v_2$ fluctuations
  $\frac{\sigma_{v_2}}{v_2} \approx
  \frac{\sigma_{\varepsilon}}{\varepsilon}$ we deduce the nonflow
  implied for several models of eccentricity fluctuations that would
  be required for consistency with $v_2\{2\}$ and $v_2\{4\}$. 
  We also
  present results on the ratio of $v_2$ to eccentricity. 
\end{abstract}


\maketitle

\vspace{0.5cm}

\section{Introduction}


In non-central heavy-ion collisions, the overlap area is almond shaped
with a long and short axis. Secondary interactions amongst the
system's constituents can convert the initial coordinate-space
anisotropy to a momentum-space anisotropy in the final
state~\cite{OllitraultSorge, firstv2, rhicv2}. In this case, the \textit{spatial}
anisotropy decreases as the system expands so that any observed
\textit{momentum} anisotropy will be most sensitive to the early phase
of the evolution before the spatial asymmetry is
smoothed~\cite{Kolb:2000sd}. Ultra-relativistic nuclear collisions
at Brookhaven National Laboratory's Relativistic Heavy Ion Collider
(RHIC)~\cite{RHIC} are studied in part to deduce whether quarks and
gluons become deconfined during the early, high energy-density phase
of these collisions. Since the azimuthal momentum-space anisotropy of
particle production is sensitive to the early phase of the collision's
evolution, observables measuring this anisotropy are especially
interesting. The azimuth angle ($\phi$) dependence of the distribution
of particle momenta can be expressed in the form of a Fourier
series~\cite{Voloshin:1994mz}: $dN/d\phi \propto 1 +
\sideset{}{_n}\sum\nolimits 2v_n\cos n\left (\phi-\Psi \right )$,
where $\Psi$ is either the reaction-plane angle defined by the
beam axis and the impact parameter vectors, or the participant plane angle
defined by the beam direction and the minor axis of the overlap
zone~\cite{Alver:2008zza,Alver:2008zza}. Fluctuations in the positions of
nucleons within the colliding nuclei can cause deviations between the
reaction plane angle and the participant plane angle and the non-sphericity of the colliding nuclei may also enhance this effect.   
When energy is deposited in the overlap region by a finite number of
collision participants, the energy density will necessarily possess a
lumpiness associated with statistical fluctuations. These fluctuations
will lead to eccentricity fluctuations which can lead to $v_2$
fluctuations. 
By definition, the eccentricity 
is maximum when calculated with respect to the participant plane. This plane shifts away from the reaction plane due to fluctuations.
It is expected that
this larger, positive definite eccentricity will drive the anisotropic
expansion thought to be responsible for
$v_2$~\cite{Alver:2008zza}. The eccentricity calculated with respect
to the participant axis is called $\varepsilon_{\mathrm{part}}$ and
the eccentricity calculated with respect to the reaction plane is
called $\varepsilon_{\mathrm{std}}$. 

The Fourier
coefficients $v_n$ can be measured and used to characterize the
azimuthal anisotropy of particle production.
Measurements of $v_{2}$ \cite{v2papers} have been taken to indicate
the matter created in collisions at
RHIC behaves like a perfect liquid with a viscosity-to-entropy ratio near
a lower bound $\eta/s>1/4\pi$ derived both from the uncertainty principle
\cite{gyulassy} and string theory \cite{Kovtun:2004de}. This
conclusion is primarily based on hydrodynamic model predictions
\cite{hydro, v2papers}. Uncertainty about the conditions at the
beginning of the hydrodynamic expansion, however, leads to large
uncertainties in the model expectations \cite{cgcecc,fklncgc}. Since
$v_{2}$ reflects the initial spatial eccentricity of the overlap
region when two nuclei collide, fluctuations of $v_{2}$ should depend on fluctuations in the initial eccentricity.  Measurements of
the system-size and energy dependence of $v_{2}$ and $v_{2}$
fluctuations are therefore useful for understanding the initial
conditions of the expansion phase of heavy-ion collisions.

Methods used to study $v_{2}$~\cite{art} are based on correlations
either among produced particles or between produced particles and
spectator neutrons detected near beam rapidity
$y_{\mathrm{beam}}$. Estimates of $v_{2}$ from produced particles can
be biased by correlations which are not related to the reaction or
participant plane (nonflow
$\delta_2 \equiv \langle\cos(2\Delta\phi)\rangle-\langle v_2^2\rangle$) and
by event-by-event fluctuations of $v_{2}$ ($\sigma_{v_2}$). Thus, an
explicit measurement of $\langle v_{2}\rangle$ would require a
measurement of nonflow and fluctuations. We also note that when the
definition of the reference frame changes, from reaction plane to
participant plane for example, each of the terms $v_2$, $\delta_2$,
and $\sigma_{v_2}$ can change. The experimentally observable
n-particle cumulants of $v_2$ (labeled $v_{2}\{2\}^2$, $v_2\{4\}^4$,
etc.) do not, however, depend on the choice of reference frame. It has
been shown~\cite{misn,opv,gmod} that the various analyses of $v_2$ based on
produced particles are related to the second and fourth $v_2$
cumulants $v_2\{2\}$ and $v_2\{4\}$ where these are related
to $v_2$, nonflow, and fluctuations in the participant plane reference frame via
\begin{linenomath}\begin{equation}
  v_2\{4\}^2 \approx \langle v_2\rangle^2 - \sigma_{v_2}^2 \label{eq:v4}
\end{equation}\end{linenomath}
and
\begin{linenomath}\begin{equation}
  v_2\{2\}^2-v_2\{4\}^2 \approx \delta_2 + 2\sigma_{v_2}^2  .  \label{eq:diff}
\end{equation}\end{linenomath}
These results arise because fluctuations decrease $v_2\{4\}$ but
increase $v_2\{2\}$ and the approximations are valid for
$\sigma_{v_2}/\langle v_2\rangle \ll 1$. We will discuss the effect of
this approximation later. In case the $v_2$ distribution is
a 2D Gaussian in the reaction plane, the
6-particle cumulant $v_2\{6\}$ and higher orders will be equal to
$v_2\{4\}$ and therefore will not add new information. Within the
accuracy of the data this has been found to be the case (\textit{i.e.}
$v_{2}\{6\}\approx v_{2}\{4\}$)~\cite{genfunction}.
In this approximation for the $v_2$ fluctuations~\cite{gmod}, $v_2\{4\}$ is equal to the
mean $v_2$ relative to the reaction plane and $\sqrt{v_2^2\{4\}+\sigma_{v_2}^2}$ is the mean $v_2$ relative to the participant plane.
We note again that $\sigma_{v_2}^2$ is not
experimentally accessible without prior knowledge about nonflow
contributions~\cite{Sorensen:2008zk}.

In this paper we present measurements of $v_2\{2\}$ and $v_2\{4\}$ in
Au+Au and Cu+Cu collisions at $\sqrt{s_{_{\mathrm{NN}}}}=200$ and 62.4
GeV. We present $v_2\{2\}^2-v_2\{4\}^2 \approx \delta_2+
2\sigma_{v_2}^2$ (called in the literature
$\sigma_{\mathrm{tot}}^{2}$) and derive from that upper limits on
$\sigma_{v_{2}}/v_2$ based on several approximations. The upper limit
assumes that $v_2$ fluctuations dominate the sum $\delta_2 +
2\sigma_{v_2}^2$. This is a robust upper limit since larger values of
$\sigma_{v_{2}}/v_2$ would require negative values of nonflow
contrary to expectations and to measurements of two-particle
correlations~\cite{correlations}.  We present model comparisons of
eccentricity fluctuations to the upper limit of
$\sigma_{v_{2}}/v_2$. Using the same data and then alternatively
assuming that eccentricity fluctuations drive $v_2$ fluctuations,
we can derive the
nonflow term required to satisfy the relationship
$v_2\{2\}^2-v_2\{4\}^2 \approx \delta_2 + 2\sigma_{v_2}^2$ for each
model. The $\delta_2$ derived in this way can be compared to
measurements of two-particle correlations~\cite{correlations} to check
the validity of the models. Finally we present the ratio of $v_2$ to
the initial eccentricity from the models.

This paper does not use the method of a global fit to a detailed
11-parameter model of two-particle correlations in relative
pseudorapidity and azimuth~\cite{2-part}. The method used here
requires no assumptions about the shape of flow fluctuations or
nonflow and instead considers Fourier harmonics of the azimuthal
distributions.

This paper is organized as follows: Section II gives the experimental
details and cuts for the data selection. Section III deals with
details about the Q-Cumulants method and the sources of systematic
errors. In Section IV, $v_2$ results used in the calculation of the
nonflow and the upper limit on $v_2$ fluctuations are discussed.
Section V shows the results for the upper limit on $v_2$ fluctuations
and their comparison with the eccentricity fluctuations, nonflow from
different models and eccentricity scaling of $v_2$ for the
eccentricity from different models.

\section{Experiment}
\label{ex}

Our data sets were collected from Au+Au and Cu+Cu collisions at
$\sqrt{s_{_{\mathrm{NN}}}} = 62.4$ and 200~GeV detected with the STAR
detector~\cite{STAR} in runs IV (2004) and V (2005). Charged particle tracking within pseudo-rapidity
$|\eta|<1$ and transverse momentum $p_T>0.15$~GeV/$c$ was performed
with the Time Projection Chamber (TPC)~\cite{tpc}. Beam-beam Counters
(BBCs) and Zero-degree Calorimeters (ZDCs) were used to trigger on events. We analyzed events from
centrality interval corresponding to 0--80\% and 0--60\% of the
hadronic interaction cross-section respectively for Au+Au and Cu+Cu
collisions. As in previous STAR analyses~\cite{scalingv2}, we define
the centrality of an event from the number of charged tracks in the
TPC having pseudorapidity $|\eta| < 0.5$~\cite{mult}. 
For the $v_2$
analysis we used charged tracks with $|\eta|<1.0$ and $0.15 < p_{T} <
2.0$ GeV/$c$. The lower $p_T$ cut is necessitated by the acceptance of
the STAR detector. We varied the upper $p_T$ cut between 1.5 and 3.0
GeV/$c$ to study the effect of this cut on the difference $v_2\{2\}^2-v_2\{4\}^2$. We found that $v_2\{2\}$ and
$v_2\{4\}$ increase by roughly 5\% (relative) when the upper $p_T$ cut
is increased from 1.5 to $3.0$~GeV/$c$, but that the difference between $v_2\{2\}^2$ and
$v_2\{4\}^2$ changes by less than 1\%. Only events with primary
vertices within 30 cm of the TPC center in the beam direction were
analyzed. The cuts used in the analysis are shown in Table~\ref{cuts}.

\begin{center}
\begin{table}
\begin{tabular} { l  r  }
\hline
\hline
cut  &value\\
\hline
$p_{T}$ & 0.15 to 2.0 GeV/$c$\\
$\eta$ & -1.0 to 1.0\\
vertex z & -30.0 cm to 30.0 cm\\
vertex x,y & -1.0 cm to 1.0 cm\\
fit points &  $>$15 \\
fit points/max. pts. &  $>$0.52 \\
dca & $<$ 3.0 cm\\
trigger &  Minbias \\
\hline
\end{tabular}
\caption{Cuts used for the selection of data. Fit points are the number of points used to fit the TPC track, and max. points are the maximum possible number for that track.}
\label{cuts}
\end{table}
\end{center}

\section{Analysis}
\label{an}

We analyzed Cu+Cu and Au+Au collisions at center of mass energies
$\sqrt{s_{_{\mathrm{NN}}}} = 62.4$ and 200~GeV to study the energy and
system-size dependence of $v_2$, nonflow and $v_2$ fluctuations. From
previous studies we found that it is not possible to use $v_2$ cumulants to disentangle
nonflow effects (correlations not related to the event plane) from
$v_{2}$ fluctuations~\cite{Sorensen:2008zk}. We have used two methods based on multi-particle azimuthal
correlations:  Q-Cumulants~\cite{qcum} for two- and four-particle cumulants to study $v_{2}\{2\}$ and $v_{2}\{4\}$, and fitting the reduced flow vector
$q=Q/\sqrt{M}$ distribution to study the multi-particle $v_2$. $Q=\sum_{j}^{M}e^{2i\phi_j}$ and $M$ is the multiplicity. The fitting of the reduced flow vector distribution is
described in more detail in Ref.~\cite{Sorensen:2008zk}. The fit
parameters described in that reference, $v_2\{\textrm{qfit}\}$ and
$\sigma_{\textrm{dyn}}^2$ (in this paper $\sigma_{\textrm{tot}}^2$),
can be related to $v_2\{2\}$ and $v_2\{4\}$.  In Appendix~\ref{comp},
we compare the q-distribution and Q-Cumulants results. Based on
simulations, we find that the q-distribution method used to study
$v_2$ by fitting the distribution of the magnitude of the reduced
flow vector to a function derived from the central limit theorem deviates more from the input values when
multiplicity is low.  For that reason, this paper presents only results
from the Q-Cumulants method.

The Q-Cumulants method allows us to calculate the cumulants without
nested loops over tracks or using generating
functions~\cite{genfunction}. For this reason it is simpler to perform. The cumulants calculated in this way also do not suffer from
interference between different harmonics since the contributions from
other harmonics are explicitly removed~\cite{qcum}. We directly
calculate the two- and four- particle azimuthal correlations
\begin{eqnarray}
\langle2\rangle_{n|n}&=&\frac{|Q_{n}|^{2}-M}{M(M-1)} \label{eq1a} \\
\langle4\rangle_{n,n|n,n}&=&\frac{|Q_{n}|^{4} +  |Q_{2n}|^{2} -2 Re[Q_{2n}Q^{*}_{n}Q^{*}_{n}]}{M(M-1)(M-2)(M-3)}\nonumber \\ 
&-&2 \frac{2(M-2)|Q_n|^2-M(M-3)}{M(M-1)(M-2)(M-3)}\, , \label{eq1b}
\end{eqnarray}
 where $M$ is the number of tracks used in the analysis and
\begin{equation}
Q_{n}=\sum_{j}^{M} e^{in\phi_j}\, .
\label{eq2a}
\end{equation}

We evaluate the terms on the right hand side of Eq.~\ref{eq1a} and Eq.~\ref{eq1b} for each event, then take the average over all events.
If one applies no further weighting, the two- and four-particle
cumulant results for $v_n$ are
 \begin{eqnarray}
v_{n}\{2\}^2&=&\langle2\rangle_{n,n} \label{eq3a} \\
v_{n}\{4\}^4&=&2 \langle2\rangle_{n,n}^{2}-\langle4\rangle_{n,n|n,n}. \label{eq3b}
\end{eqnarray}

It was also proposed to use weights for each event within a particular
centrality class based on the number of combinations of tracks for each
event~\cite{qcum}. 
This weighting was proposed as a method to reduce the dependence of
the results on multiplicity. We find however, that the application of
number-of-combinations weights makes the $v_2\{2\}$ and $v_2\{4\}$
results more dependent on the width of the multiplicity bins used to
define centrality in our analysis. Using number-of-combination weights
along with centrality bins defined by number of charged particles will
lead to results that are weighted more heavily towards the higher
multiplicity side of the bins and that effect will be stronger for
four-particle correlations than for two-particle correlations. We also
confirmed with simulations that without weights, the Q-Cumulant
results for $v_2\{2\}$ and $v_2\{4\}$ agree better with simulation
inputs than when weights are applied. In this paper, we report results
without weights according to Eqs.~\ref{eq1a} through \ref{eq3b}. This
method is different from that used in Ref.~\cite{alice}.

The systematic uncertainties on our results were estimated by
evaluating our results from two different time periods in the run, by
varying the selection criteria on the tracks (specifically the
distance of closest approach of the track to the primary vertex or
DCA), from the Q-Cumulants acceptance correction terms, and by varying
the $p_T$ upper limit for tracks between 1.5, 2.0, and 3.0 GeV/c.
Decreasing the DCA cut and increasing the upper $p_T$ cut both
increase the average $p_T$ of the analyzed tracks. This leads to an
increase in $v_2\{2\}$ and $v_2\{4\}$ (not considered a systematic
error for those data) but we find that the difference between
$v_2\{2\}^2$ and $v_2\{4\}^2$ is nearly unchanged. This implies that
the error on $v_2\{2\}^2-v_2\{4\}^2$ due to the exact upper and lower
$p_T$ ranges used is small. We found no difference between the two run
periods analyzed. The acceptance correction applied in the analysis
changes the 200 GeV Au+Au Q-Cumulants $v_2\{4\}$ results by less than
1\% for all centralities while the $v_2\{2\}$ results change by less
than 1\% for all centralities except the 0-5\% bin where they change
by 4\%, and the 5-10\% bin where they change by 2\%. Statistical and
systematic errors are shown on all results. The systematic errors are
shown as narrow lines with wide caps and statistical errors are shown
as thick lines with narrow caps. In many cases statistical
errors are smaller than the marker size and therefore not
visible.

\begin{figure*}[thb]
\centering\mbox{
\includegraphics[width=0.5\textwidth]{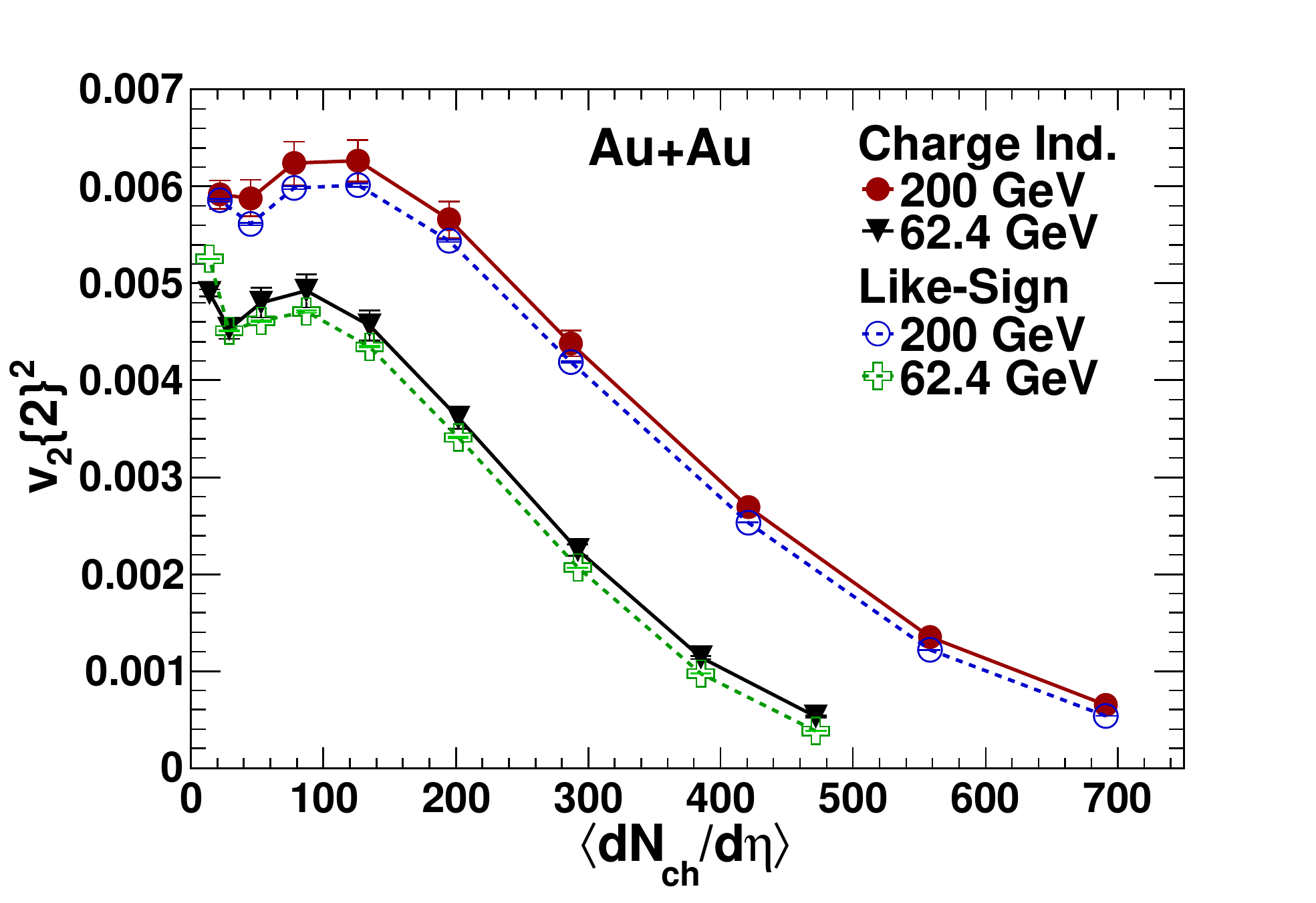}
\includegraphics[width=0.5\textwidth]{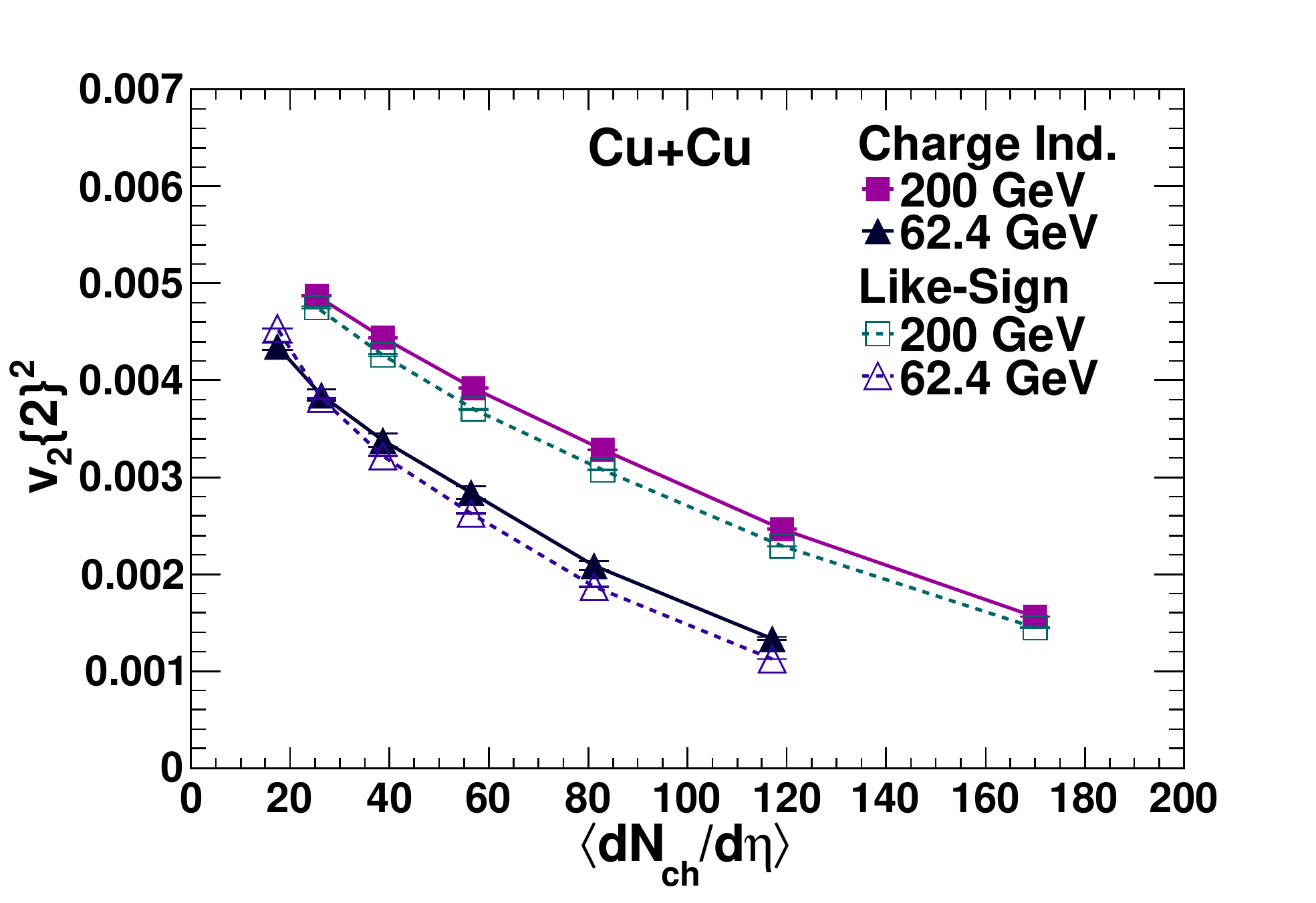}}
\caption{ Left: The two-particle cumulant $v_2\{2\}^{2}$ for Au+Au
  collisions at 200 and 62.4 GeV. Results are shown with like-sign
  combinations (LS) and charge-independent results (CI) for
  $0.15<p_T<2.0$~GeV/$c$. Right: The same as the left but for Cu+Cu
  collisions. The systematic errors are shown as thin lines with wide
  caps at the ends and statistical errors are shown as thick
  lines with small caps at the end. Statistical and systematic
  errors are very small. } \label{fig1b} \end{figure*}

\section{Results}
\label{re}

\begin{figure}[htbp]
\centering\mbox{
\includegraphics[width=0.5\textwidth]{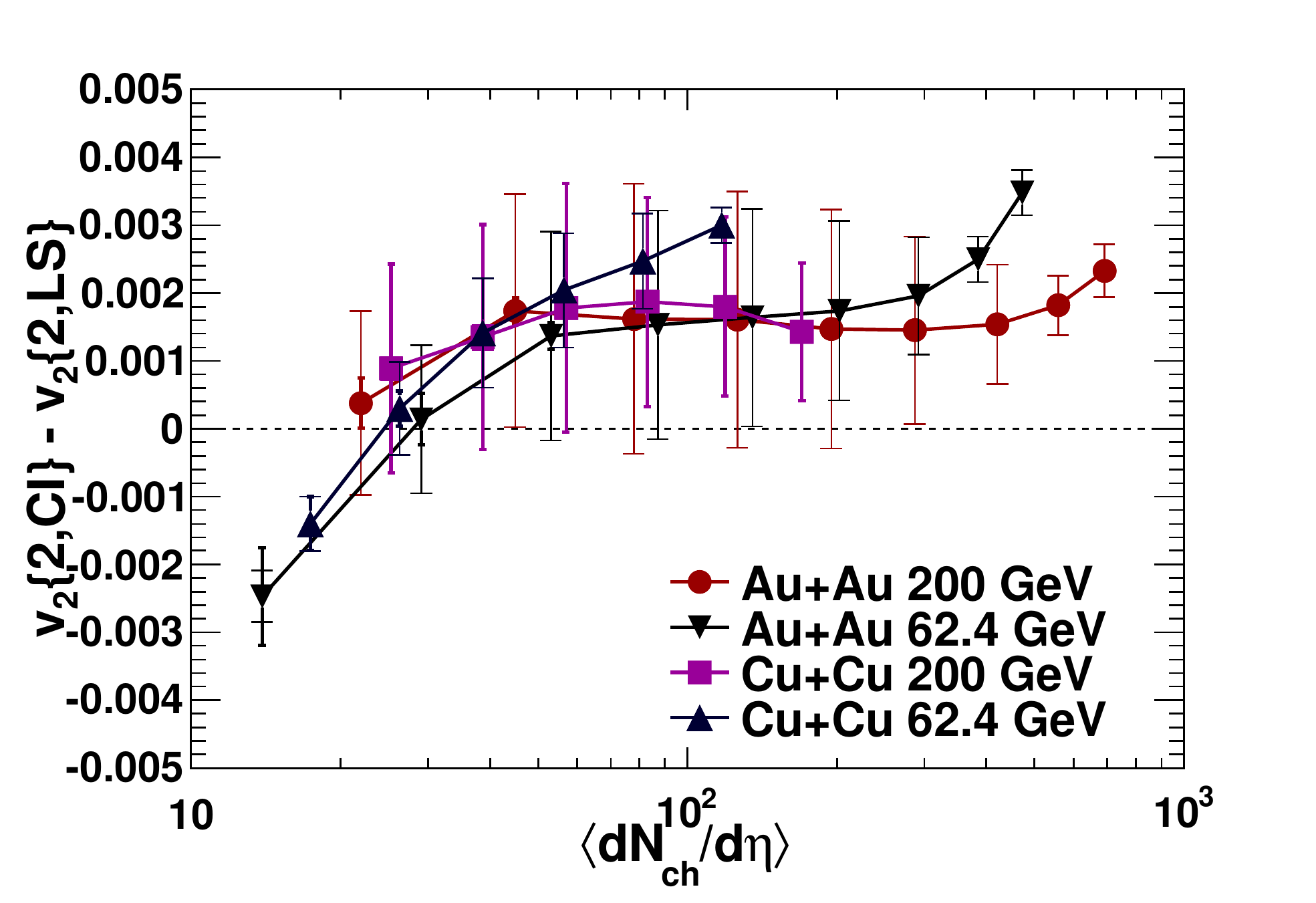}}
\caption{ The difference of 
  charge-independent (CI) $v_2\{2\}$ and like-sign (LS) $v_2\{2\}$ for Au+Au and
  Cu+Cu collisions at 200 and 62.4 GeV vs. the log of $\langle dN_{\mathrm{ch}}/d\eta\rangle$. The statistical errors are smaller than the marker size and not visible for most of the data. } \label{fig2c}
\end{figure}

In this paper we present our results as a function of the average
charged particle multiplicity density $\langle
dN_{\mathrm{ch}}/d\eta\rangle$ within a given centrality interval.
Table~\ref{cent} in Appendix~\ref{mo} provides estimates of the number
of participating nucleons $N_{\mathrm{part}}$ and $\langle
dN_{\mathrm{ch}}/d\eta\rangle$ for the centrality intervals used in
this analysis. Figure~\ref{fig1b} (left) shows $v_2\{2\}^{2}$ for 200
and 62.4 GeV Au+Au collisions for charged tracks with
$0.15<p_T<2.0$~GeV/$c$. The analysis is carried out using either all
combinations of particles, independent of charge (CI), or using only
like-sign pairs (LS). When comparing the LS and CI results, we note
that the LS results are systematically lower than the CI results for
all centralities except the most peripheral bin. This behavior might
be related to nonflow since many known nonflow effects lead to
correlations preferentially between opposite sign particles;
\textit{e.g.} neutral resonances decay into opposite sign particles
and jet fragments tend to be charge ordered~\cite{Abreu:1997wx}. The
LS results therefore typically contain smaller nonflow correlations.
 Bose-Einstein correlations between identical particles, on the other hand, can lead to larger nonflow for LS than for CI. 
Figure~\ref{fig1b} (right) shows the CI and LS results for Cu+Cu
collisions at $\sqrt{s_{\mathrm{NN}}}=$200 and 62.4 GeV. The same
trends hold with the LS results lower than the CI results.
 %
Figure~\ref{fig2c} shows the difference of  CI $v_2\{2\}$ and LS $v_2\{2\}$
for Au+Au and Cu+Cu collisions at 200 and 62.4 GeV. This difference shows a dependence on energy only for central Au+Au collisions. In the lowest multiplicity data, CI $v_2\{2\}$ becomes smaller than LS $v_2\{2\}$, consistent with expectations from Bose-Einstein correlations.



\begin{figure*}[thb]
\centering\mbox{
\includegraphics[width=0.5\textwidth]{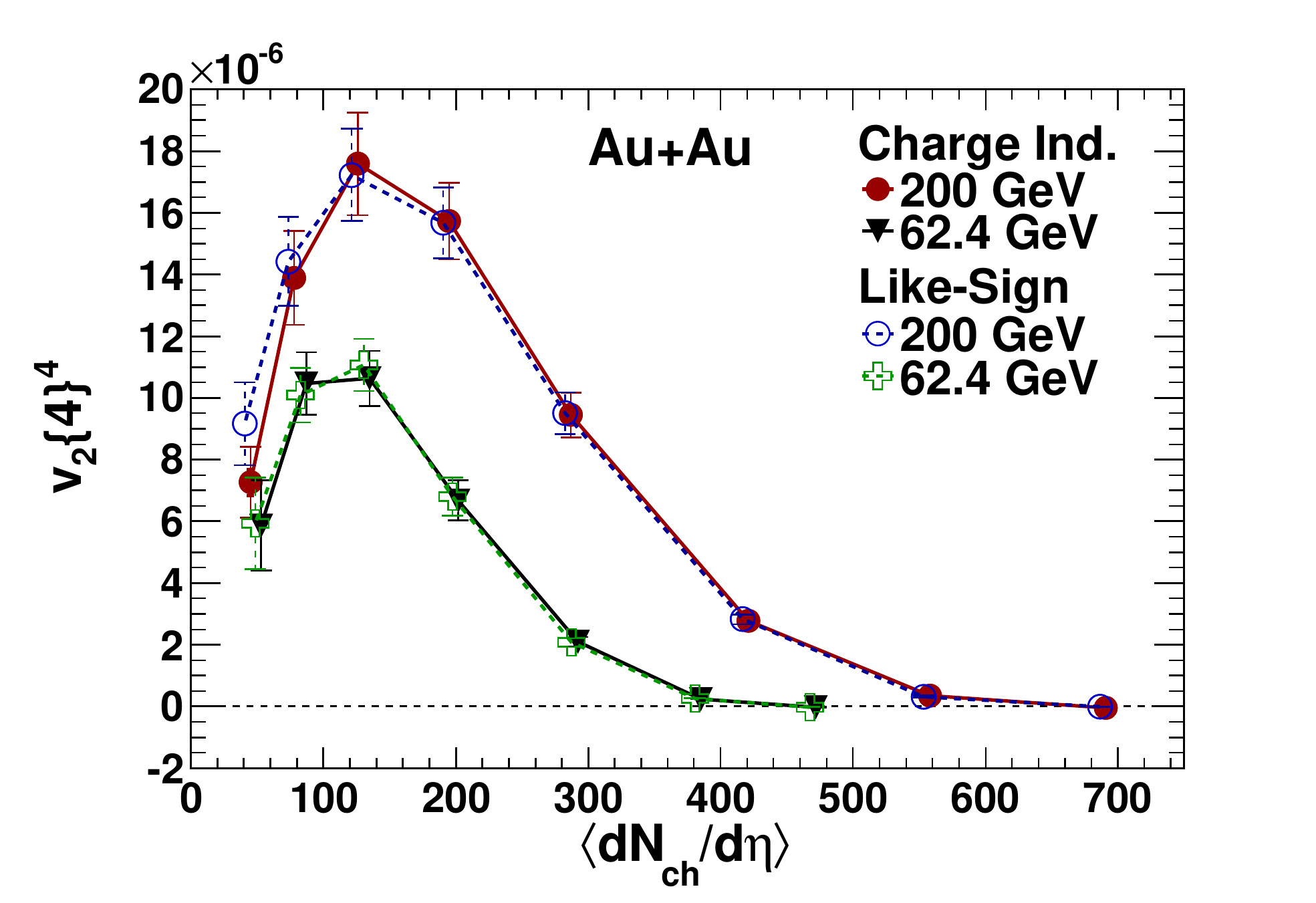}
\includegraphics[width=0.5\textwidth]{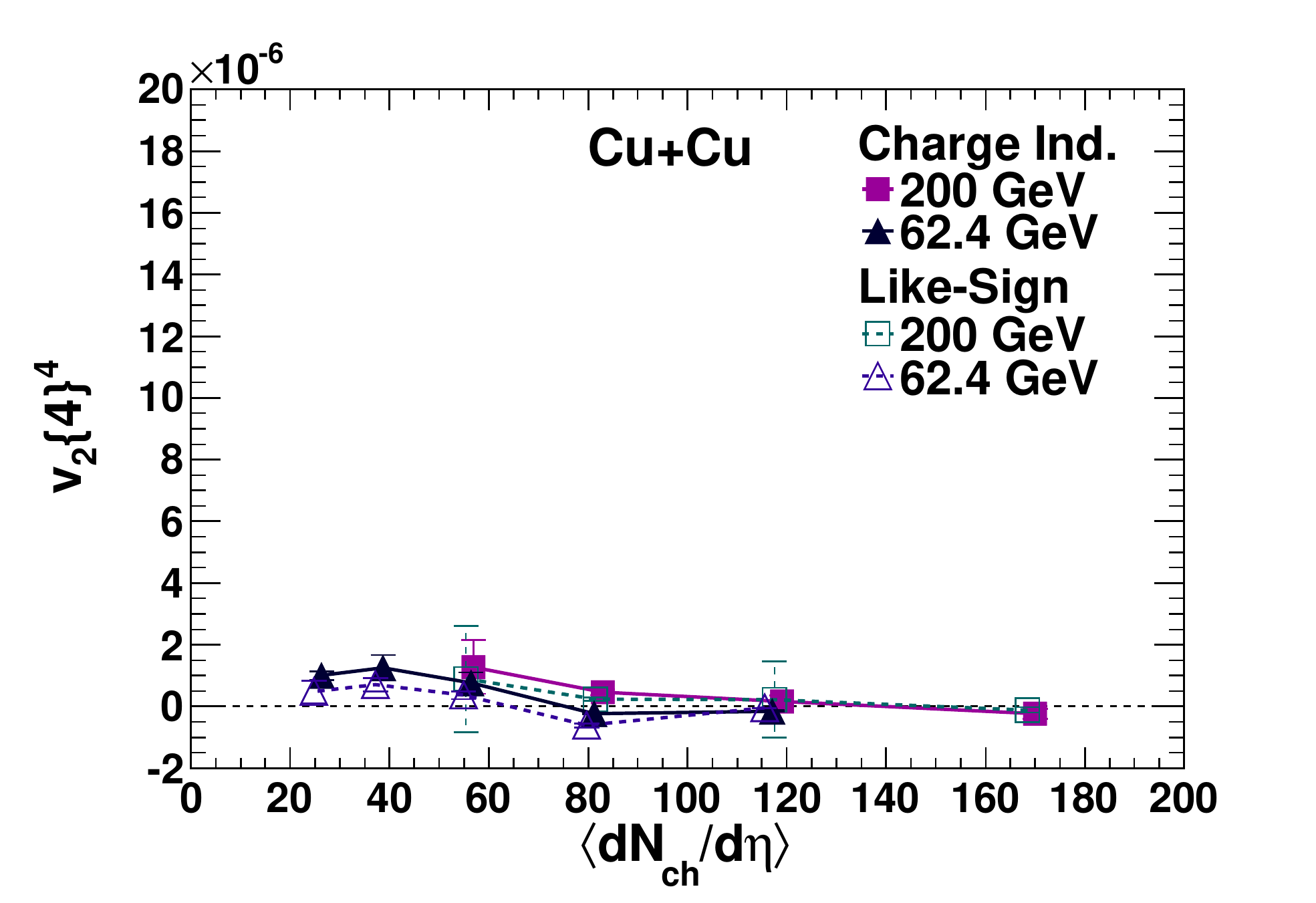}}
\caption{ Left: The LS and CI four-particle cumulant $v_2\{4\}^{4}$
  for Au+Au collisions at 200 and 62.4 GeV for $0.15<p_T<2.0$~GeV/$c$.
  The systematic errors are shown as narrow lines with wide caps at
  the end and statistical errors are shown as thick lines with narrow
  caps at the end. Statistical errors are not visible for most of the
  points. Right: The LS and CI four-particle cumulant $v_2\{4\}^{4}$
  for Cu+Cu collisions at 200 and 62.4 GeV for $0.15<p_T<2.0$~GeV/$c$.
  The most central points (two points for Cu+Cu 62.4 GeV) gives
  $v_{2}\{4\}^{4}<0$ for all the data sets.
  The negative values are probably due to large fluctuations in
  agreement with Eq.~(1). These may include contributions from impact
  parameter spread and finite multiplicity bin width. } \label{fig2a}
\end{figure*}

Figure~\ref{fig2a} shows the four-particle cumulant $v_2\{4\}^{4}$ for
Au+Au (left) and Cu+Cu (right) collisions at 200 and 62.4 GeV. In the
case of $v_2\{4\}^{4}$, no differences are detected between LS and CI
results. This suggests that nonflow correlations are suppressed as
expected in the four-particle cumulant results. Any nonflow source
leading to fewer than four correlated particles will not contribute to
$v_2\{4\}^{4}$. In addition, while any nonflow for $v_2\{2\}^2$ is
suppressed only by $1/M$, any nonflow correlations between four or
more particles will still be suppressed by a combinatorial factor of
$(M-1)(M-2)(M-3)$. $v_2\{4\}^{4}$ shows slightly negative values for the more
central events for Au+Au and Cu+Cu collisions at 200 and 62.4 GeV.
$v_2\{4\}^{4}$ is allowed to take on negative values. These may be
associated with $v_2$ fluctuations larger than those expected from
eccentricity fluctuations alone. In this case however, the second or
fourth roots of $v_2\{4\}^{4}$ cannot be defined. For this reason,
those points are not included in the analysis of $v_2\{2\}^2 -
v_2\{4\}^{2}$. All results are reported in the data
tables~\cite{onlinetabs}. It had been observed from simulations that
the measurement of $v_2\{4\}$ using the Q-Cumulants method deviates
from input for the most peripheral collisions. Also, the LS $v_2\{4\}$
data appears to scatter for mean charged particle multiplicity density
$\langle dN_{\mathrm{ch}}/d\eta\rangle <$ 26. Therefore, no data
points are used for comparison with models for $\langle
dN_{\mathrm{ch}}/d\eta\rangle <$ 26.

\begin{figure*}[htb]
\centering\mbox{
\includegraphics[width=0.5\textwidth]{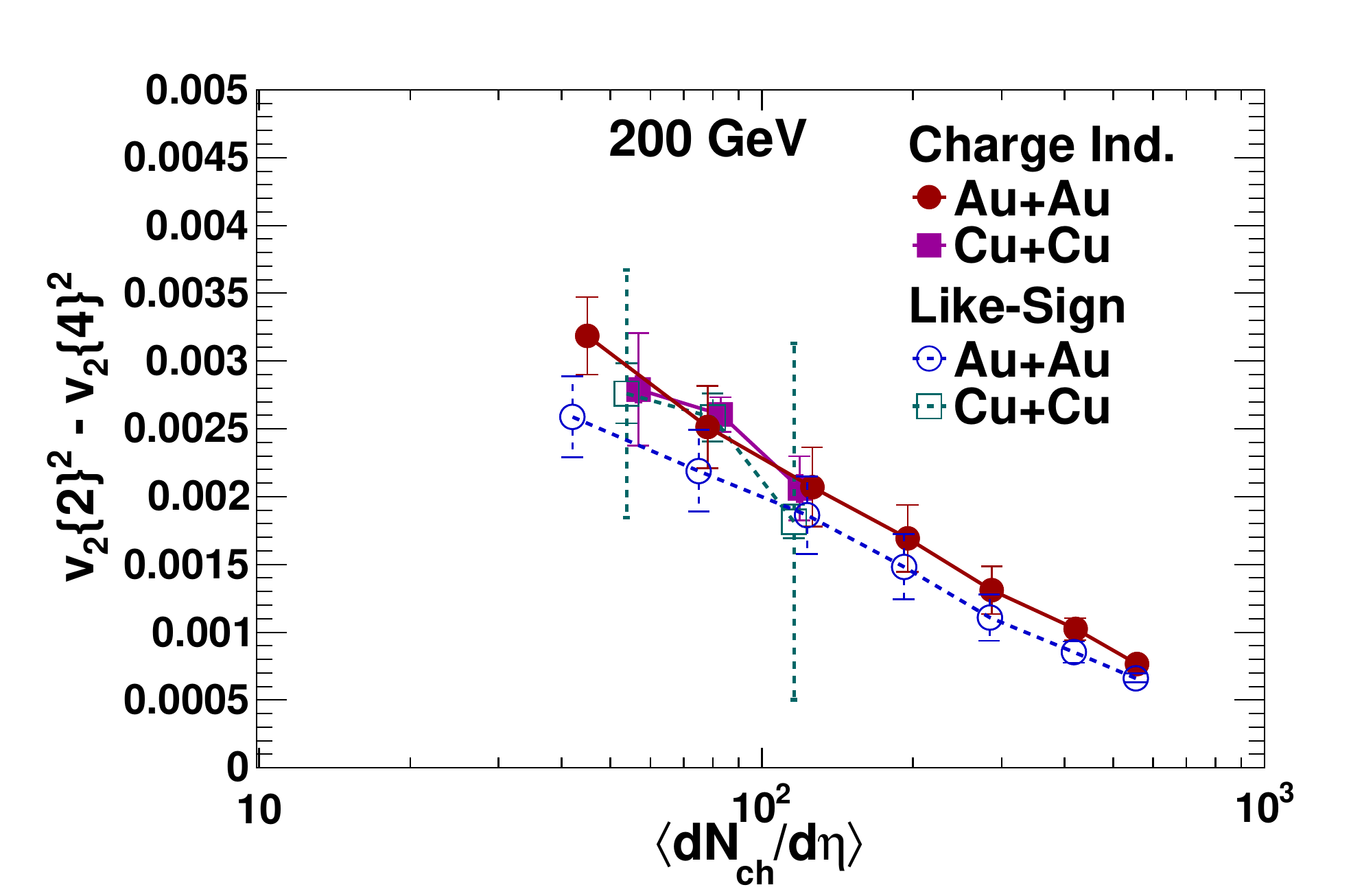}
\includegraphics[width=0.5\textwidth]{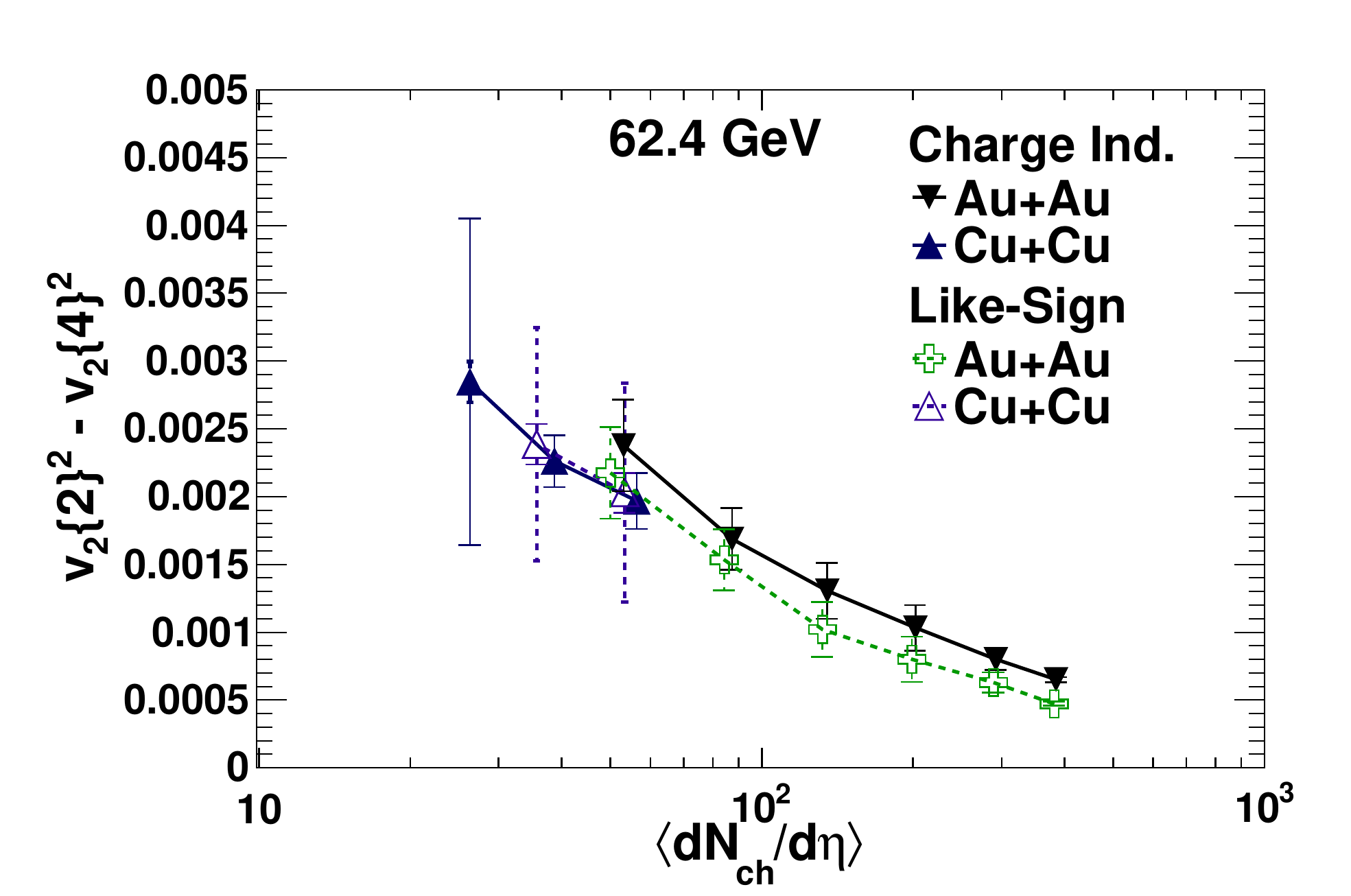}}
\caption{Left: The difference between $v_2\{2\}^2$ and $v_2\{4\}^2$
  for 200 GeV Au+Au and Cu+Cu collisions for both LS and CI
  combinations. Right: The difference between $v_2\{2\}^2$ and
  $v_2\{4\}^2$ for 62.4~GeV
  Au+Au and Cu+Cu collisions for both LS and CI combinations. The statistical and
  systematic errors are shown as in previous figures. }
\label{fig3a} 
\end{figure*}

Figure~\ref{fig3a} shows $v_2\{2\}^2-v_2\{4\}^2$ for Au+Au and Cu+Cu
collisions at 200 GeV (left) and 62.4 GeV (right) for both LS and CI.
The difference between $v_2\{2\}^2$ and $v_2\{4\}^2$ is of interest
because it is related to nonflow $\delta_2$ and $v_2$ fluctuations:
\begin{linenomath}\begin{equation}
v_2\{2\}^2-v_2\{4\}^2 \approx \delta_2 + 2\sigma_{v_2}^{2} \equiv \sigma_{\mathrm{tot}}^2.
\label{eq:sigtot}
\end{equation}\end{linenomath}
This difference can be taken as an approximate upper limit on nonflow
$\delta_2$. We estimate that the approximation in Eq.~\ref{eq:sigtot}
which assumes $\langle v_2\rangle$ is much larger than the second,
third and fourth moments of $v_2$ is accurate to within 30\% for these
data sets. We arrive at this estimate by assuming $v_{2} \propto
\varepsilon_{\mathrm{part}}$ and then using our Monte Carlo Glauber
model to calculate
$(\varepsilon_{\mathrm{part}}\{2\}^2-\varepsilon_{\mathrm{part}}\{4\}^2)/2\sigma_{\varepsilon_{\mathrm{part}}}^2$.
If the approximation in Eq.~\ref{eq:sigtot} is accurate, this ratio
should be unity. We find that for the centralities considered here,
the ratio is within 30\% of unity. Below, where we compare our data to
eccentricity models, a significant fraction should cancel since the
approximation applies to both the data and the models.
The difference $v_2\{2\}^2-v_2\{4\}^2$ increases with beam energy and
decreases with increasing mean multiplicity. The contribution from
nonflow typically scales as 1/${\it \langle dN_{\mathrm{ch}}/d\eta
  \rangle}$ if the number of clusters scales with ${\it \langle
  dN_{\mathrm{ch}}/d\eta \rangle}$ and the number of particles per
cluster is constant. A $1/N_{\mathrm{part}}$ dependence is also
expected for $\sigma_{v_2}^{2}$ from eccentricity fluctuations. The
energy dependence can come from either an increase in nonflow
correlations with energy and/or an increase in $v_2$ fluctuations with
energy. The LS results are systematically lower than the CI results
for all but the lowest multiplicities, consistent with a nonflow
contribution to the CI $v_2\{2\}$ results which is reduced for the LS
$v_2\{2\}$ results. In the model comparisons that follow, we will use
the LS results to compare our results to three eccentricity models.

\section{Data and Eccentricity Models}

We compare our $v_2\{2\}$ and $v_2\{4\}$ results characterizing the distribution of $v_2$, to equivalent measures characterizing the eccentricity distributions of three models.  
The models are a
Monte-Carlo Glauber model with nucleons as participants (MCG-N), a Monte-Carlo Glauber model with quarks as participants
(MCG-Q), and a CGC based Monte-Carlo model
(fKLN-CGC). The models are described in more detail in
Appendix~\ref{mo}. Another analysis of models has been published in Ref.~\cite{Heinz}. The non-sphericity of the Au nuclei has been neglected in eccentricity calculations for the models because non-sphericity only affects the most central collisions which are not used in the comparison of data with models~\cite{deformedau}. 

\subsection{Upper Limit on Relative Fluctuations}

We would like to compare our data to models for eccentricity fluctuations
by comparing $\sigma_{v_2}/v_2$ to
$\sigma_{\varepsilon}/\varepsilon$. We can not uniquely determine the
value of $\sigma_{v_2}$ from the two- and four-particle cumulant data
however, since $v_2\{2\}^2-v_2\{4\}^2 \approx
\delta_2+2\sigma_{v_2}^2$. We can however derive an upper limit on the
ratio $\sigma_{v_2}/v_2$ by setting $\delta_2=0$. This amounts to
assuming the difference between the two- and four-particle cumulant is
dominated by $v_2$ fluctuations and that $\delta_2$ cannot be
negative. Although negative nonflow values can easily be
generated from resonance decays in specific kinematic regions, we
consider the case that the total nonflow should become negative
highly unlikely and contradictory to studies of the nonflow effect. 
The quantity
\begin{linenomath}\begin{equation}
    R_{v\mathrm (2-4)} = \sqrt{\frac{v_2\{2\}^2-v_2\{4\}^2}{v_2\{2\}^2+v_2\{4\}^2}}
    \label{vupper}
  \end{equation}\end{linenomath}
then becomes an upper limit to the ratio $\sigma_{v_2}/\langle
v_2\rangle$ where, in the case that $v_2$ fluctuations are dominated
by eccentricity fluctuations, $\langle v_2\rangle$ is the average
$v_2$ relative to the participant axis~\cite{gmod}. Additional
fluctuations from another source will lead to a contribution to the
difference between $v_2\{2\}$ and $v_2\{4\}$ not related to the
eccentricity fluctuations that relate the reaction plane and the
participant plane. In the following figures, we compare the ratio
$R_{v\mathrm (2-4)}$ for the like-sign results to the ratio
\begin{linenomath}\begin{equation}  
  R_{\mathrm \varepsilon(2-4)} = \sqrt{\frac{\varepsilon\{2\}^2-\varepsilon\{4\}^2}{\varepsilon\{2\}^2+
\varepsilon\{4\}^2}}
\label{eupper}
\end{equation}\end{linenomath} 
for the three eccentricity models described in Appendix~\ref{mo},
where $\varepsilon\{2\}$ and $\varepsilon\{4\}$ are the second and
fourth cumulants for $\varepsilon_{\mathrm{part}}$. Since higher
moments (skewness and kurtosis) of the distribution of $v_2$ or
$\varepsilon_2$ contribute to Eqs. \ref{vupper} and \ref{eupper} it is
important to compare the same quantities from data and the
eccentricity models. For small $\varepsilon$ Eq.~\ref{eupper} becomes
$\sigma_{\varepsilon}/\langle\varepsilon\rangle$.

\begin{figure*}[htbp]
\centering\mbox{
\includegraphics[width=0.5\textwidth]{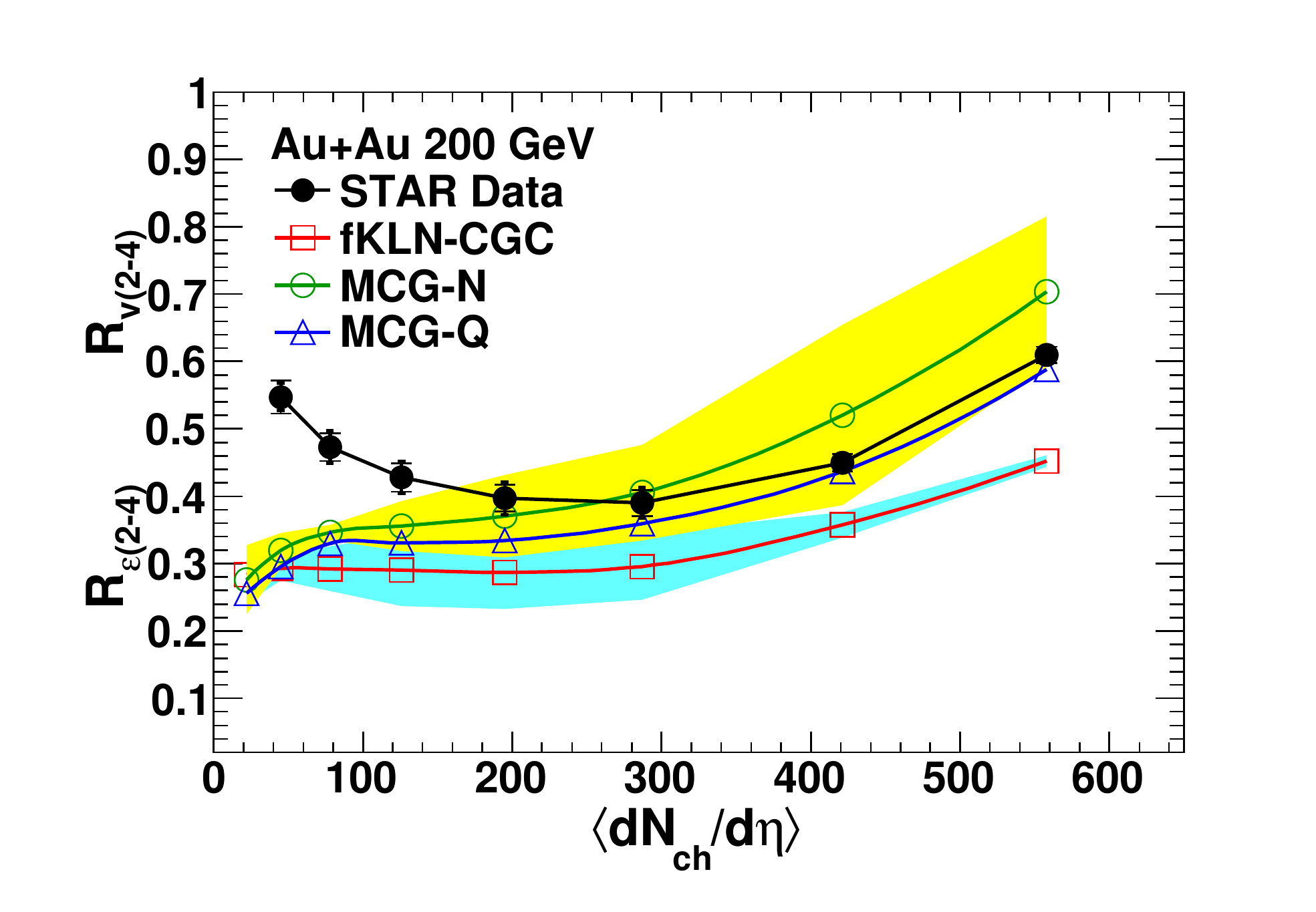}
\includegraphics[width=0.5\textwidth]{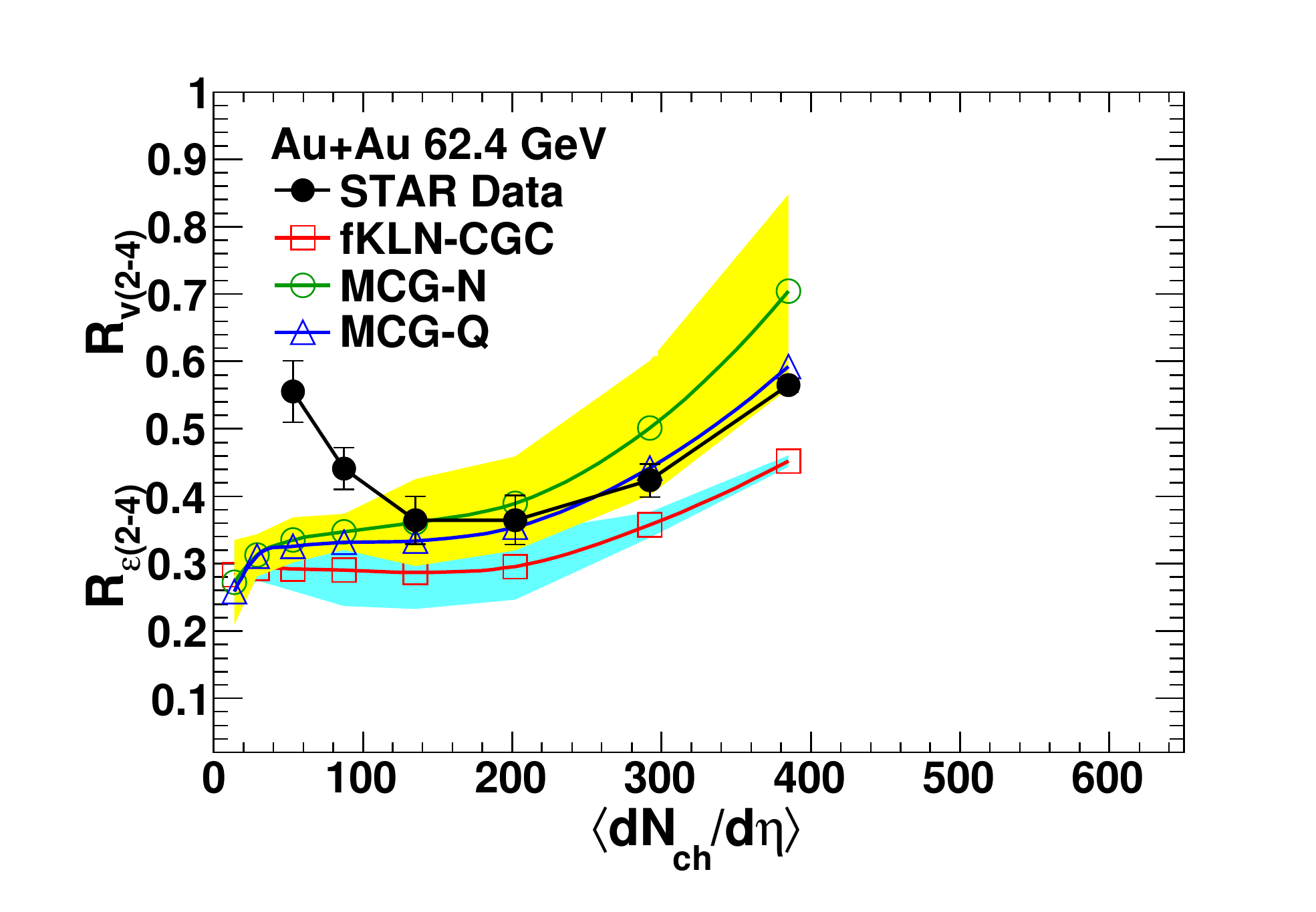}}
\caption{ The upper limit on $\sigma_{v_2}/\langle v_{2}\rangle$ for
  200 GeV (left) and 62.4 GeV (right) Au+Au collisions from
  Eq.~(\ref{vupper})
  compared to $\sigma_{\varepsilon}/\varepsilon$ from
  Eq.~(\ref{eupper}) for three different models. The upper limit is
  found using the LS results for $v_{2}\{2\}$. Data are from the range
  $0.15<p_T<2.0$~GeV/$c$. The shaded bands reflect the
  uncertainties on the models which are dominated by uncertainty on
  the distribution of nucleons inside the nucleus. The uncertainty is
  only shown for the MCG-N and fKLN-CGC models. The uncertainty
  on the MCG-Q model is the same as for the MCG-N model but is not shown
  for the visual clarity.}
\label{fig10}
\end{figure*}

Figure~\ref{fig10} shows $R_{v\mathrm (2-4)}$ vs mean charged hadron
multiplicity for 200 GeV (left) and 62.4 GeV (right) Au+Au data. The
LS $v_2\{2\}$ results are used to reduce nonflow. The data is compared
to the same quantity for the three different models. The shaded bands
show the uncertainties on the models that arise primarily from the
uncertainty in the Woods-Saxon parameters used to describe the nuclei.
The error is correlated between Monte-Carlo models and for clarity is
only plotted on the MCG-Q and fKLN-CGC models. The centrality in the
models is defined using multiplicity so that the model calculations
include bin-width effects and impact parameter fluctuations similar to
data. In as much as the models correctly model the multiplicity, by
defining centrality in the models the same way that it is defined in
data, both the model and the data will have the same impact parameter
fluctuations.

In peripheral collisions ($\langle dN_{\mathrm{ch}}/d\eta\rangle<
150$), data exceeds the eccentricity models substantially. This is not
surprising since we expect a significant contribution from nonflow in
this region. The central value for the ratio from the MCG-N model
rises with increasing centrality and then overshoots the upper limit in the most
central collisions. Given the errors indicated by the yellow band
however, the MCG-N model could still be consistent with the upper
limit. The MCG-Q model approaches the upper limit in central
collisions but never exceeds it. The fKLN-CGC model has the smallest
values and is well below the upper limit throughout the entire
centrality range. Notice that in the models, the more constituents,
the smaller the fluctuations.


In Fig.~\ref{fig10} (right), the 62.4 GeV Au+Au data are compared to
models. Data points are only reported where $v_2\{4\}^4$ is positive.
At this lower energy, peripheral data is again above the models. The
central value for the MCG-N model again overshoots the upper limit for
central and mid-central collisions while the MCG-Q model appears to
just reach the upper limit for the most central data point. The
uncertainty on the geometry of the Au nucleus again however makes it
impossible to rule out any of the models in this comparison. The
fKLN-CGC model lies below the upper limit for the entire range. The
fact that the MCG-N and MCG-Q models reach and in some cases exceed
the upper limit means that for those models to be correct, nonflow
would have to be small or perhaps even negative. Nonflow can be
negative from resonance decay, but is not likely. The lower energy
data therefore provides a very useful test of the models and results
from the beam energy scan at RHIC promise to provide even better
constraints~\cite{bes}.

\begin{figure}[htbp]
\centering\mbox{
\includegraphics[width=0.5\textwidth]{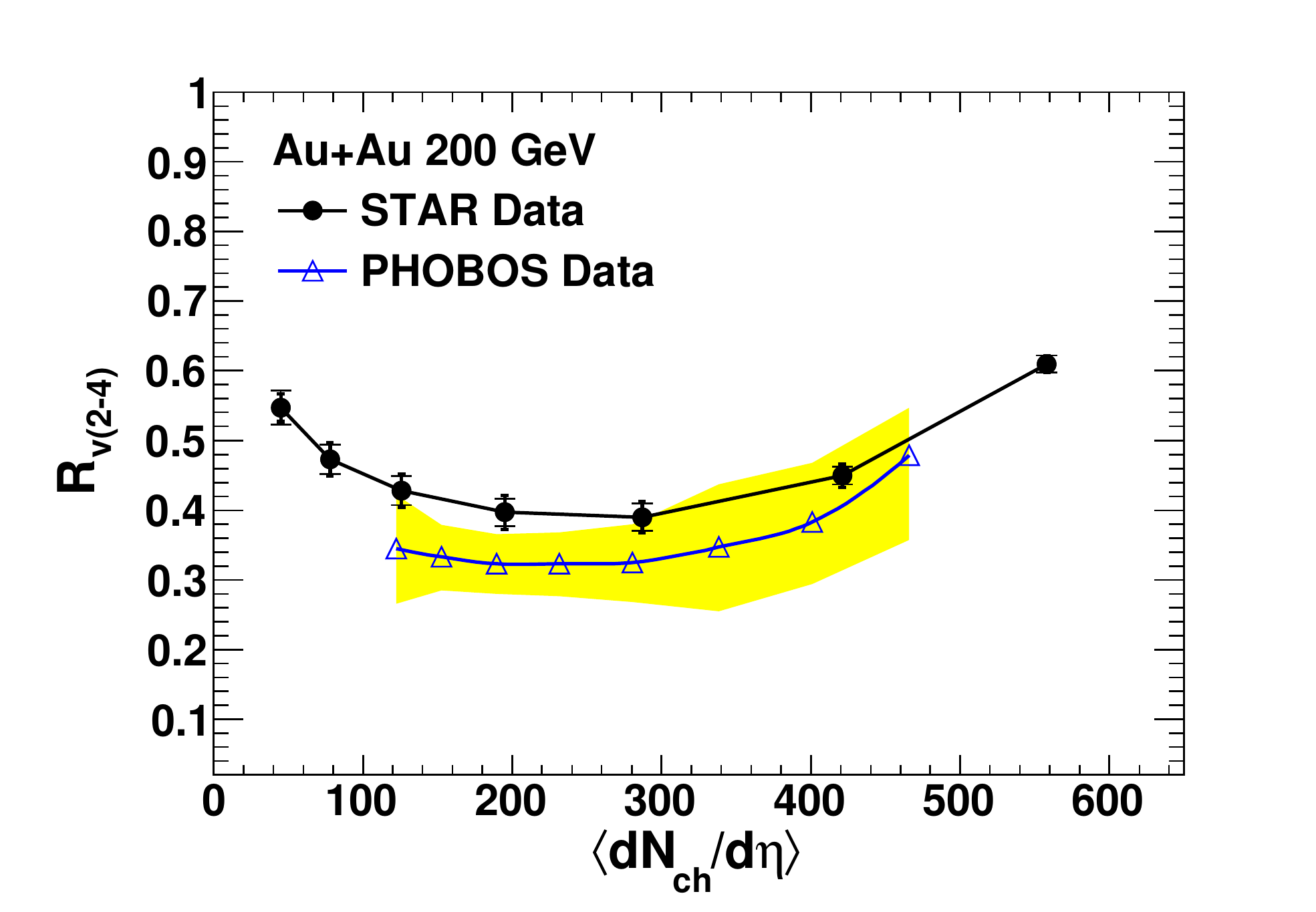}}
\caption{The STAR data compared to PHOBOS data~\cite{Alver:2010rt} on
 $\sigma_{v_2}/\langle v_2\rangle$ with $\delta_2$ for $\Delta\eta>2$
 taken to be zero (see Fig. 6 from Ref.~\cite{Alver:2010rt}). The
 shaded band shows the errors quoted from Ref.~\cite{Alver:2010rt}.}
\label{fig10b} \end{figure}

Figure~\ref{fig10b} shows the STAR 200 GeV Au+Au data on the upper
limit for $\sigma_{v_2}/\langle v_2\rangle$ compared to the PHOBOS
results reported in Ref.~\cite{Alver:2010rt} under their assumption
that $\delta_2$ is zero for $\Delta\eta>2$ (see the reference for
details). The PHOBOS results are for all charged particles while the
STAR results are for LS pairs only. PHOBOS has subtracted narrow
$\Delta\eta$ correlations by fitting $v_2(\eta_1)v_2(\eta_2)$ and
removing the narrow diagonal peak corresponding to small-$\Delta\eta$
nonflow correlations. This may explain why the PHOBOS results are
slightly below the STAR upper limits derived from LS $v_2\{2\}$,
suggesting that there may be some residual nonflow in our LS results.
We also note however, that the analysis procedures in this paper and
in the PHOBOS paper are quite
different.

\begin{figure*}[htbp]
\centering\mbox{
\includegraphics[width=0.5\textwidth]{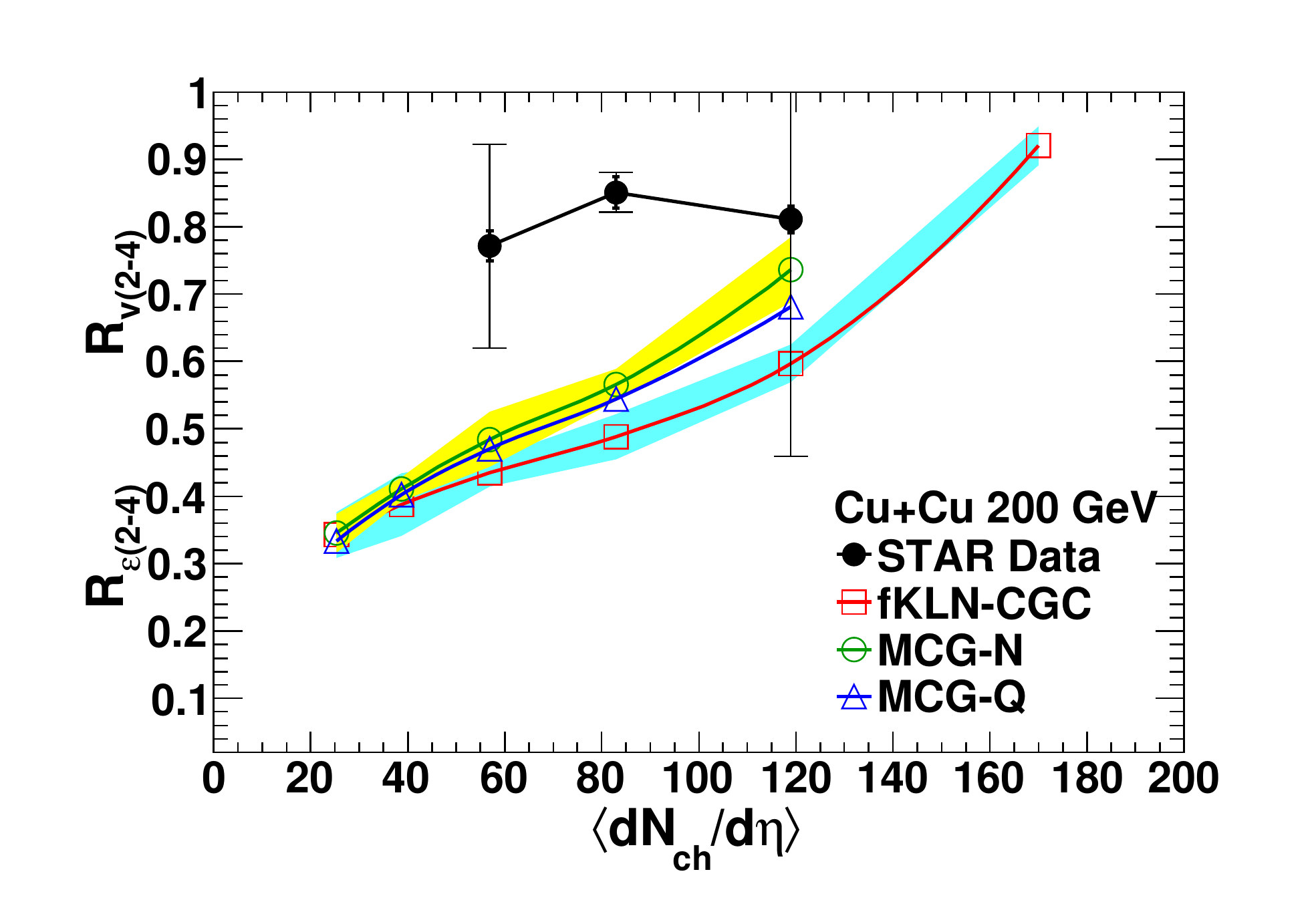}
\includegraphics[width=0.5\textwidth]{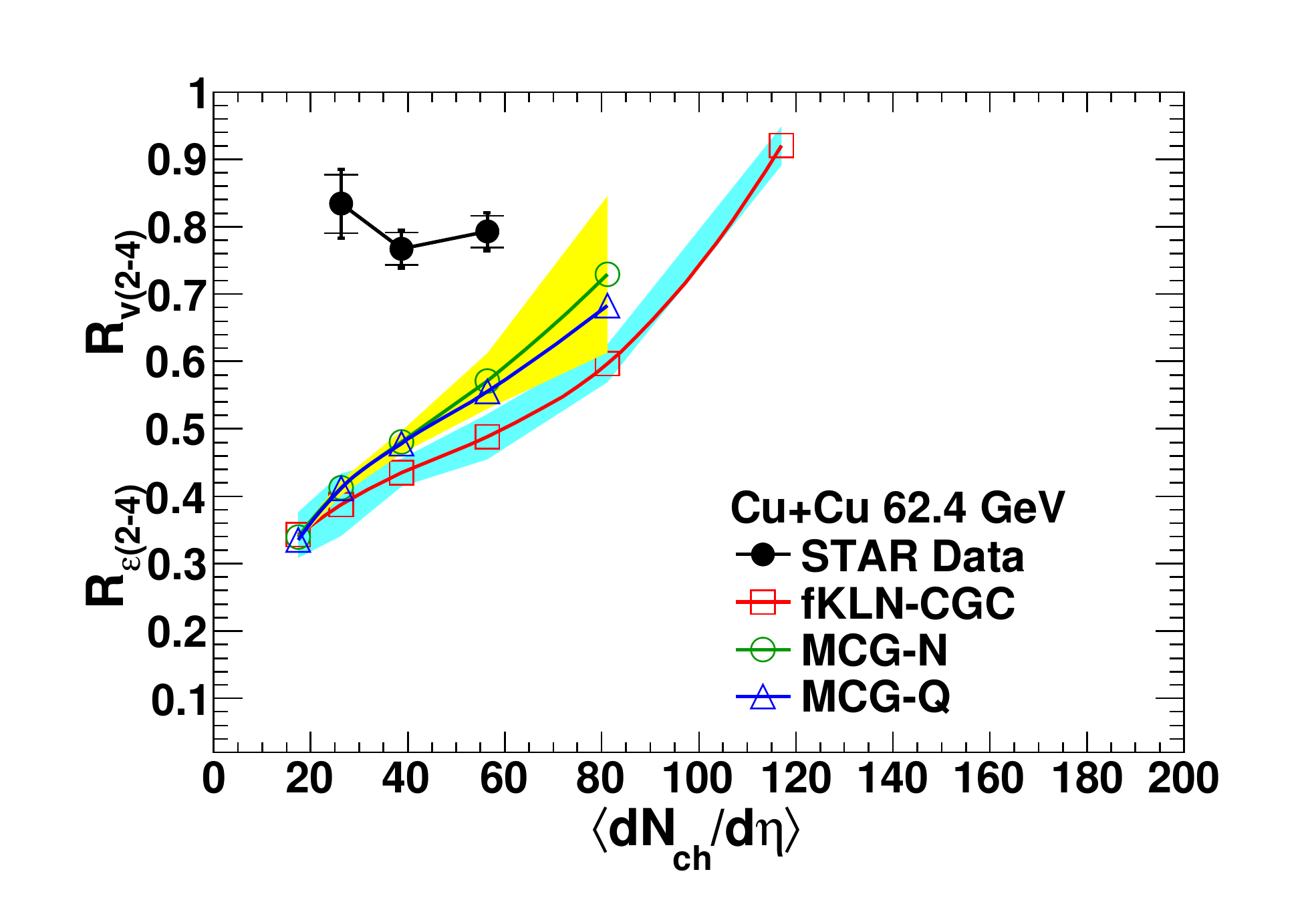}}
\caption{The upper limit on $\sigma_{v_2}/\langle v_{2}\rangle$ for
  200 GeV (left) and 62.4 GeV (right) Cu+Cu collisions from
  Eq.~(\ref{vupper}) compared to $\sigma_{\varepsilon}/\varepsilon$
  from Eq.~(\ref{eupper}) for three different models. } \label{fig12}
\end{figure*}

Figure~\ref{fig12} shows the upper limits and models for Cu+Cu
collisions at 200 and 62.4 GeV (respectively left and right). Data
points are only reported where $v_2\{4\}^4$ is positive. The upper
limit on fluctuations for Cu+Cu collisions are larger than for Au+Au
and lie near unity. All the models fall below the upper limit and
differences between the models are small. This is likely due to the
large multiplicity fluctuations for smaller systems in the models
which masks the other physical differences between the models. The
large Cu+Cu results do not provide constraint on the models.


\subsection{Nonflow}

An alternative way to look at the $v_2\{2\}^2$ and $v_2\{4\}^2$ data
is to
apply the assumption
that $v_2$ fluctuations are dominated by the initial spatial
eccentricity fluctuations
\begin{linenomath}\begin{equation}
    \sigma_{v_2} \approx \langle v_2\rangle \frac{\sigma_{\varepsilon}}{\varepsilon}
    \label{eq:fluct}
  \end{equation}\end{linenomath}
and then derive the nonflow $\delta_2$ that would be implied by each
eccentricity model. In Eq.~\ref{eq:fluct}, $\langle v_{2}\rangle $ is
not directly observable. Then we can calculate the value of $\delta_2$
that would be needed to satisfy the following equation:
\begin{linenomath}\begin{equation}
  v_2\{2\}^2-v_2\{4\}^2 \approx \delta_2 + 2\sigma_{v_2}^2.
\end{equation}\end{linenomath}
Recalling from Eq.~\ref{eq:v4} and~\ref{eq:diff} that
$v_2\{2\}^2+v_2\{4\}^2 \approx \delta_2 + 2\langle v_2\rangle^2$, we
derive the following expression for $\delta_2$:
\begin{linenomath}
\begin{equation}
\delta_2 \approx v_2\{2\}^2-v_2\{4\}^2\left(\frac{\varepsilon^2+\sigma_{\varepsilon}^2}{\varepsilon^2-\sigma_{\varepsilon}^2}\right).
\end{equation}
\end{linenomath}
which only depends on the directly observable cumulants and quantities
obtained from models. Since a model dependence exists, the $\delta_2$
values are not measurements of $\delta_2$ but instead provide an
alternative consistency check for the models.
These values can be compared to other measurements of nonflow
correlations such as the already measured two-particle
correlations~\cite{2-part}. This is an important test for the models,
since a complete model of heavy ion collisions should be able to
predict multiple observables at once. The interpretation, however, of
the structures in two-particle correlations such as the
ridge~\cite{2-part} is in flux. In particular, the nonflow
correlations from jets are inferred from two-particle correlations vs
$\Delta\eta$ and $\Delta\phi$ after subtracting a $\Delta\eta$
independent $v_2^2$ term. This approximation may not be valid for
reasons discussed recently in the literature~\cite{torque}. Given the
current state of understanding, in this paper we do not make a direct
comparison of the nonflow correlations inferred from this analysis to
those inferred from two-particle correlations.

In the absence of new physics, the term $\delta_2$ will vary with
event multiplicity as $1/M$. This is because, in the case that high
multiplicity events are a linear superposition of lower multiplicity
events, the numerator in the mean grows as $M$ while the denominator
grows as the number of pairs $M(M-1)/2$. To cancel out the
combinatorial $1/M$ dependence we scale $\delta_2$ by the number of
mean charged hadrons within $|\eta|<0.5$. A variation of $\langle
\frac{dN_{\mathrm{ch}}}{d\eta}\rangle\delta_2$ with multiplicity
implies a non-trivial change in the physics.

\begin{figure*}[htbp]
\centering\mbox{
\includegraphics[width=1.0\textwidth]{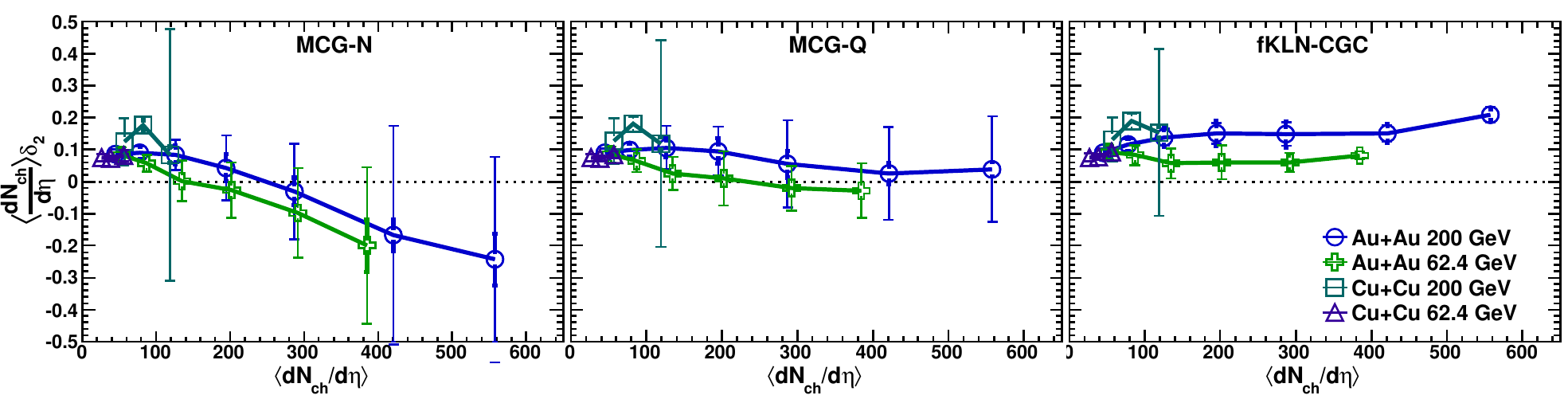}}
\caption{The mean multiplicity scaled nonflow $(\delta_{2})$ found by
  assuming Eq.~\ref{eq:fluct} and using the three different models for
  $\sigma_{\varepsilon}^2/\varepsilon^2$: MCG-N (left), MCG-Q
  (middle), and fKLN-CGC (right). The systematic errors are shown as thin lines with wide
  caps and statistical errors are shown as thick lines
  with narrow caps. Statistical errors
  are not visible for most of the points.}
 \label{fig17}
\end{figure*}

Figure~\ref{fig17} (left) shows the like-sign
$\langle\frac{dN_{\mathrm{ch}}}{d\eta}\rangle\delta_2$ that is
required if the Monte-Carlo Glauber model with nucleon participants
gives the correct description of the eccentricity fluctuations and
eccentricity fluctuations dominate $v_2$ fluctuations. The nonflow is
larger at 200 GeV than at 62.4 GeV. Within errors,
$\langle\frac{dN_{\mathrm{ch}}}{d\eta}\rangle\delta_2$ is the same in
Cu+Cu collisions and Au+Au collisions at the same energies and event
multiplicities. The value of
$\langle\frac{dN_{\mathrm{ch}}}{d\eta}\rangle\delta_2$ required by
this model seems to fall with centrality and the central value becomes
negative for the most central Au+Au collisions. The errors shown in
the figure are dominated by the systematic errors on the MCG-N model
which are highly correlated from point to point (they depend on the
parameters for the geometric description of a Au nucleus). As such,
while the most central data point is still consistent with zero, the
dropping trend with centrality is significant for most of the range
allowed for describing Au nucleus. For this model of eccentricity
fluctuations to be valid, the nonflow in central Au+Au collisions
would have to be near zero or negative. This appears to contradict
measurements showing significant near-side two-particle
correlations~\cite{2-part,Adams:2006tj}. In case other sources besides
eccentricity fluctuations contribute to $v_2$ fluctuations, the
inferred nonflow would need to become even smaller. While the
uncertainties arising from the geometric description of the Au nucleus
preclude a definitive statement, it appears likely that the MCG-N
model over-predicts the ratio of eccentricity fluctuations to the mean
eccentricity.


Figure~\ref{fig17} (middle) shows the
$\langle\frac{dN_{\mathrm{ch}}}{d\eta}\rangle\delta_2$ required if the
Monte-Carlo Glauber model with constituent quark participants gives
the correct description of the eccentricity fluctuations and if
eccentricity fluctuations dominate $v_2$ fluctuations. Within errors,
$\langle\frac{dN_{\mathrm{ch}}}{d\eta}\rangle\delta_2$ is the same in
Cu+Cu collisions and Au+Au collisions at the same energies and event
multiplicities. The smaller relative fluctuations for the constituent
quark participant model means this model would be consistent with
larger nonflow values than the nucleon participant model. The required
nonflow values are essentially positive at all measured
multiplicities. This means this model has a better chance of
accommodating the near-side two-particle correlations observed in
data.


Figure~\ref{fig17} (right) shows
$\langle\frac{dN_{\mathrm{ch}}}{d\eta}\rangle\delta_2$ derived using
the fKLN-CGC Monte-Carlo model. This model has a larger average
eccentricity and smaller eccentricity fluctuations leading to the
smallest relative fluctuations of the three models. The mean
multiplicity scaled nonflow again is larger for 200 GeV collisions
than 62.4 GeV collisions and Cu+Cu collisions seem to have the same
nonflow values as Au+Au when they are compared at the same mean
multiplicity. The multiplicity scaled nonflow implied by the fKLN-CGC
eccentricity model increases slightly or remains flat with centrality.
CGC models for the initial conditions of heavy-ion collisions have
also been invoked to try to explain the near-side correlations
observed in data~\cite{Dumitruadded}. This analysis adds information
from four-particle correlations not accessible through measurements of
a two-particle correlation function. It remains to be seen if a
consistent determination of two- and four-particle cumulants related
to $v_2$, $v_2$ fluctuations and nonflow can be derived from a CGC
model with radially boosted flux tubes.

\subsection{Eccentricity Scaling of $v_2$}

Now we show the ratio $\langle
v_2\rangle/\langle\varepsilon\rangle$ for the three models of
eccentricity. In the case that $v_2 \propto \varepsilon$, then
$\langle v_2\rangle/\langle\varepsilon\rangle$ in the reaction plane
is given by $v_2\{4\}/\varepsilon\{4\}$~\cite{Bhalerao:2006tp}
($\varepsilon\{4\}$ is the fourth cumulant defined in Appendix~\ref{mo}). In
Fig.~\ref{fig20} we plot
$v_2\{4\}/\varepsilon\{4\}$ vs mean multiplicity for Au+Au and Cu+Cu
collisions at 200 and 62.4 GeV. 
When plotted vs mean multiplicity,  $v_2\propto\varepsilon$ from all systems and energies seem to fall on top of each other. The fKLN-CGC model
displays a saturation with $v_2\propto\varepsilon$. The Monte Carlo
Glauber model with Nucleon participants shows the steepest increase of
$v_2/\varepsilon$ while the constituent quark model is intermediate
between the sharp rise of the nucleon participant model and the
saturation of the fKLN-CGC model. The approximation that $v_2 \propto
\varepsilon$ is strongly violated for the nucleon participant model. This also implies that $v_2\{4\}/\varepsilon\{4\} = \langle
v_2\rangle/\langle\varepsilon\rangle$ may be broken since that equality holds only when $v_2\propto\varepsilon$. The violation of $v_2 \propto
\varepsilon$ also implies that if the nucleon participant model is the
correct eccentricity model, then the collisions at RHIC may be far
from the ideal hydrodynamic limit. The fKLN-CGC model and constituent
quark model imply $v_2$ saturates or nearly saturates in central Au+Au
collisions, consistent with a nearly perfect liquid behavior.

\begin{figure*}[htbp]
\centering\mbox{
\includegraphics[width=1.0\textwidth]{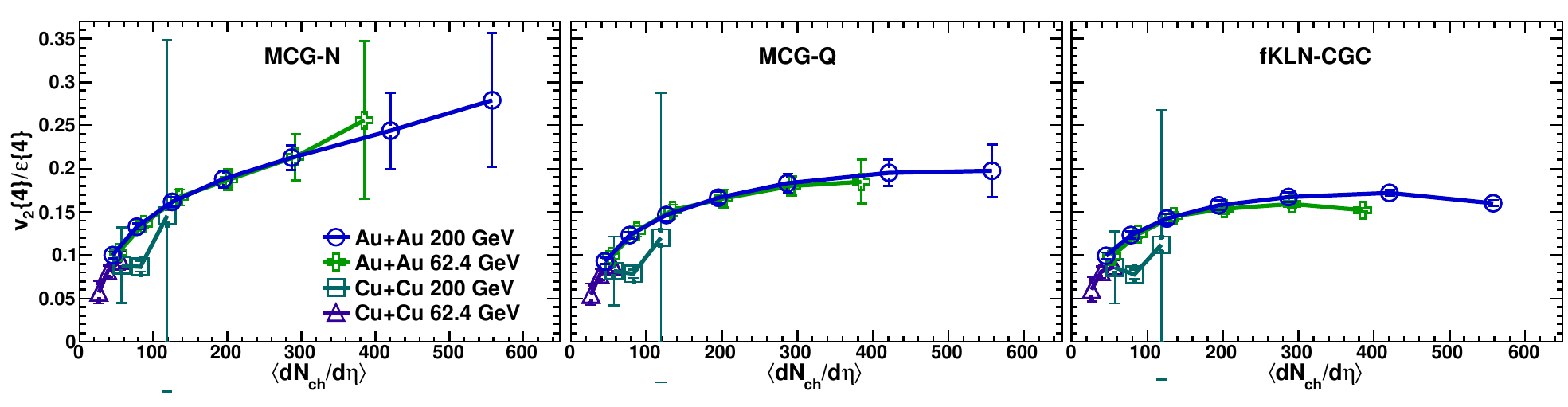}}
\caption{The eccentricity scaled $v_2$ for 200 and 62.4 GeV Au+Au and
  Cu+Cu collisions with eccentricity taken from the MCG-N (left),
  MCG-Q (middle), or fKLN-CGC (right) model. The statistical and
  systematic errors are shown as in previous figures. Statistical
  errors are not visible for most of the points.}
   \label{fig20}
\end{figure*}



\section{Conclusions}
\label{co}

We presented STAR measurements of two- and four-particle $v_2$
cumulants ($v_2\{2\}$ and $v_2\{4\}$) for Au+Au and Cu+Cu collisions
at $\sqrt{s_{_{\mathrm {NN}}}}$ = 200 and 62.4 GeV along with the difference
$v_2\{2\}^2 - v_2\{4\}^2 \approx \delta_2 + 2\sigma_{v_2} \equiv
\sigma_{\mathrm{tot}}^2$ for charge-independent and like-sign
combinations of particles. $v_2\{4\}^4$ shows negative values for the most central collisions for all the data sets. 
The difference $v_2\{2\}^2 - v_2\{4\}^2$
increases with beam energy for both Cu+Cu and Au+Au collisions. For a
given $\sqrt{s_{{\mathrm {NN}}}}$ and mean charged particle multiplicity, $v_2\{2\}^2 -
v_2\{4\}^2$ values are the same in Cu+Cu and Au+Au collisions within
errors. Although the value of $v_2$ fluctuations can not be uniquely
determined in this way, $v_{2}\{2\}$ and $v_2\{4\}$ were used to place
an upper limit on the ratio $\sigma_{v_2}/v_{2}$. The eccentricity
fluctuations from the MCG-N model are largest, rising above the upper limit from
data for central Au+Au collisions, but the MCG-Q and fKLN-CGC
eccentricity models fall within the presented limit. To further
investigate the models we calculated the value of the nonflow
$\delta_2$ implied by the models for eccentricity fluctuations under
the assumption that
$\sigma_{v_2}/v_2=\sigma_{\varepsilon}/\varepsilon$. The nonflow
values implied by the fKLN-CGC model are larger than those from either
of the Monte Carlo Glauber models. The nonflow implied by the
fluctuations in the MCG models becomes zero or negative for central
Au+Au collisions. This analysis challenges theoretical models of
heavy-ion collisions to describe all features of the data including
$v_2$, $v_2$ fluctuations, and the various correlations data. We presented
$v_2/\varepsilon$ for the three different eccentricity models and
found that the fKLN-CGC model for eccentricity leads to a saturation
of $v_2/\varepsilon$ for Au+Au collisions with $\langle\frac{dN_{\mathrm{ch}}}{d\eta}\rangle>300$ while $v_2/\varepsilon$ is rising at all
centralities when the MCG-N model is used for $\varepsilon$. 
Assuming fKLN-CGC to describe the initial state eccentricity, the saturation of $v_2/\varepsilon$ provides support for a nearly perfect hydrodynamic behavior for heavy-ion collisions at RHIC.

\section*{Acknowledgments}
We thank the RHIC Operations Group and RCF at BNL, and the NERSC
Center at LBNL for their support. This work was supported in part by
the Offices of NP and HEP within the U.S. DOE Office of Science; the
U.S. NSF; the BMBF of Germany; CNRS/IN2P3, RA, RPL, and EMN of France;
EPSRC of the United Kingdom; FAPESP of Brazil; the Russian Ministry of
Science and Technology; the Ministry of Education and the NNSFC of
China; IRP and GA of the Czech Republic, FOM of the Netherlands, DAE,
DST and CSIR of the Government of India; Swiss NSF; the Polish State
Committee for Scientific Research; VEGA of Slovakia, and the Korea
Sci. \& Eng. Foundation.

\appendix
\section{Q-Cumulants vs Fitting q-distributions}
\label{comp}

\begin{figure}[htbp]
\centering\mbox{
\includegraphics[width=0.45\textwidth]{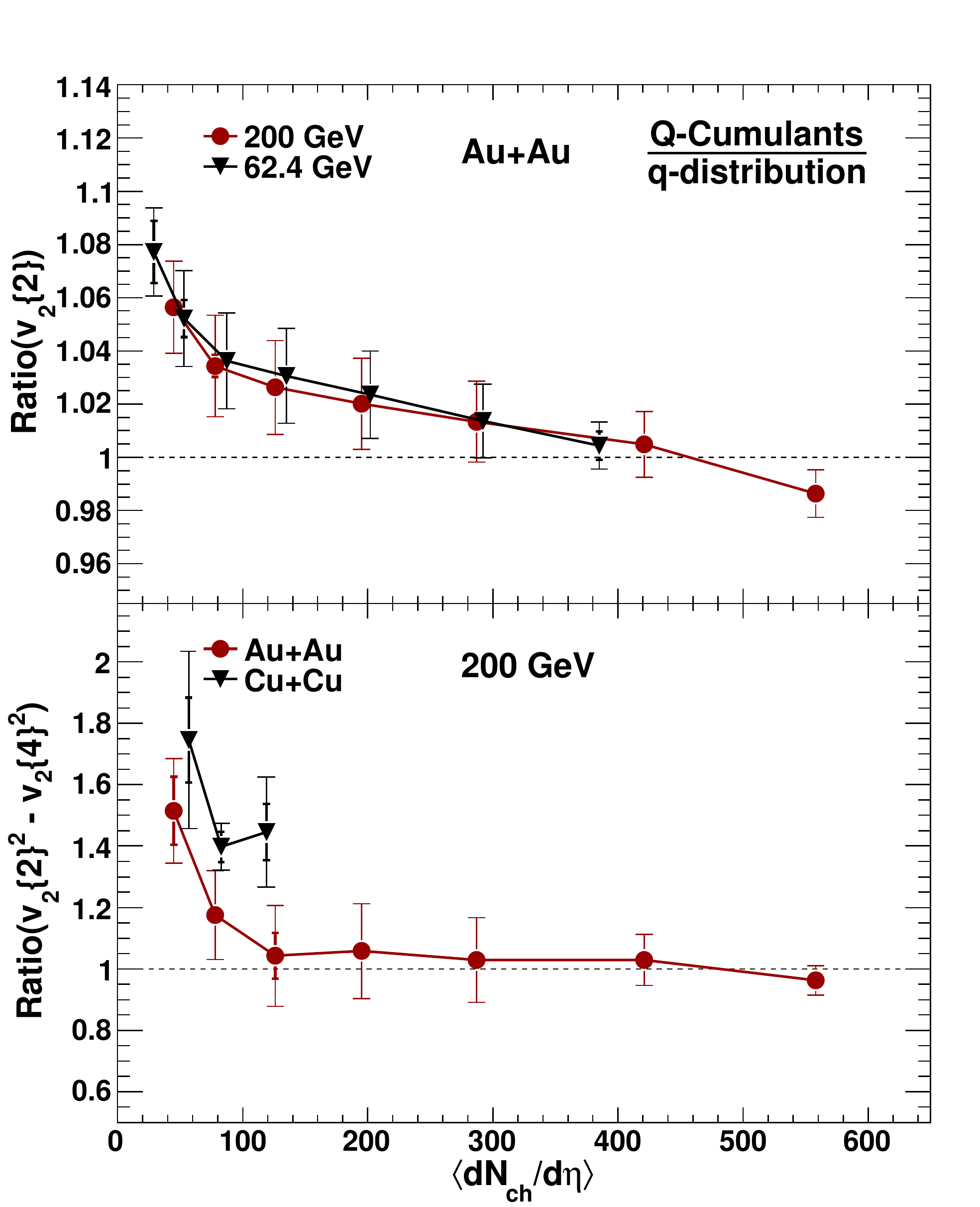}}
\caption{ Top panel: the ratio of the two-particle cumulant $v_2\{2\}$ for Au+Au
  collisions at 200 and 62.4 GeV evaluated using the q-distribution method and
  the Q-Cumulants method. Both results are calculated for combinations
  of particles independent of their charge (CI). 
Bottom panel: The ratio of the q-distribution and the Q-Cumulants method
  results for $v_2\{2\}^2 - v_2\{4\}^2$ (CI) for 200 GeV
  Au+Au and Cu+Cu collisions. In both panels, systematic errors
 are shown as thin lines with wide caps and statistical errors are shown
 as thick lines with narrow caps. 
 }
  \label{fig1a}
\end{figure}

The fitting of the reduced flow vector distribution is described in
more detail in Ref.~\cite{Sorensen:2008zk}. The fit parameters
described in that reference can be transformed to $v_2\{2,\textrm{qfit}\}^2
\equiv v_2\{\textrm{qfit}\}^2+\sigma_{\textrm{tot}}^2$ and
$v_2\{4,\textrm{qfit}\}^2=v_2\{\textrm{qfit}\}^2$ where $v_2\{2,\textrm{qfit}\}$ and
$v_2\{4,\textrm{qfit}\}$ are the two- and four-particle cumulants determined
from the q-distribution which can be compared to other determinations
of $v_2\{2\}$ and $v_2\{4\}$.  In Fig.~\ref{fig1a} (top) we show the ratio
of $v_2\{2\}$ determined from the q-distribution analysis and the Q-Cumulants analysis.

Deviations
between the q-distribution and Q-Cumulants results can be seen when the
multiplicity of the event is smaller with the q-distribution results
being smaller than the Q-Cumulants results. These deviations can be
traced to the break-down of the large N approximation required when
fitting the q-distribution. An attempt is made to correct for this
break-down which brings the results closer together but the deviations
are still significant for multiplicities below 150. The correction is
carried out by adjusting the q-distribution data before
it is fit. The correction is derived by taking the ratio of the
expected and observed q-distribution from simulated data. Although the
correction extends the apparent validity of the q-distribution
analysis to lower multiplicities, we find that the q-distribution
analysis is less reliable than the Q-Cumulants analysis.

Figure~\ref{fig1a} (bottom) shows the ratio of the quantity
$v_2\{2\}^2-v_2\{4\}^2$ from the q-distribution fits over the same
from the Q-Cumulants analysis. Data are from 200 GeV Au+Au and Cu+Cu
collisions.  
The two methods produce significantly different results
for $v_2\{2\}^2-v_2\{4\}^2$ with the difference most pronounced in
Cu+Cu and peripheral Au+Au collisions. The q-distribution gives
smaller values. This is related to the large N approximation required
in the fitting procedure for the q-distribution. When multiplicity is
low, the tails of the q-distribution cannot be populated. We find that
this leads to a narrowing of the observed distribution relative to the
fit function and the width of the distribution determines
$v_2\{2\}^2-v_2\{4\}^2$. The q-distribution fits therefore
underestimate $v_2\{2\}^2-v_2\{4\}^2$, so we use the results from the
Q-Cumulants calculation in this paper.


\section{Three Eccentricity Models}
\label{mo}
 
We use three Monte-Carlo models to study eccentricity and eccentricity
fluctuations. The first two are Glauber models which either treat
nucleons as participants or constituent quarks within the nucleons as
participants (MCG-N and MCG-Q respectively). The third model is the
factorized Kharzeev, Levin, and Nardi Color Glass Condensate model
(fKLN-CGC)~\cite{fklncgc}. The input parameters used for the
Woods-Saxon distribution of nucleons are in Table II. The Au nuclei
have been assumed spherical for the eccentricity calculations. A 0.4
fm exclusion radius is used in the calculations so that nucleons do
not overlap in coordinate space. The MCG-N model is described
elsewhere~\cite{MCGref,Alver:2008zza} and is used to calculate the
$N_{\mathrm{part}}$ and $N_{\mathrm{bin}}$ values in Table~\ref{cent}.
For the MCG-Q model, we first distribute nucleons
inside a nucleus according to a Woods-Saxon distribution with
parameters taken from~\cite{devries}, then we distribute three
constituent quarks inside each nucleon according to another Woods-Saxon
distribution where the radius of the nucleon is taken to be 0.63 fm
and the surface width is 0.08 fm. The results were not very sensitive
to variations of these parameters within a reasonable range. One might
consider a Gaussian for the quarks instead of a Woods-Saxon
distribution. The Woods-Saxon distribution gives a more flat-topped
distribution but the calculated eccentricity and eccentricity
fluctuations are not highly sensitive to the exact distribution. The
main feature of the MCG-Q model is that the potential number of
participants increases by a factor of three and there are large
correlations between participants because the quarks are confined
within the nucleons.

\begin{table}[htdp]
\begin{center}
\begin{tabular}{c c c}
\hline
\hline 
Parameter/System&$^{197}$Au~+~$^{197}$Au & $^{63}$Cu~+~$^{63}$Cu  \\
\hline 
R & 6.38  $\pm$0.06 fm  & 4.218 $\pm$ 0.014 fm\\ 
a &  0.535 $\pm$ 0.027 fm& 0.596 $\pm$ 0.005 fm\\
\hline
\hline
\end{tabular}
\label{models}
\end{center}
\caption{Input parameters for  Woods-Saxon distribution in Monte-Carlo Models.}
\end{table}%

The Woods-Saxon parameters from~\cite{devries} are based on
measurements of electron scattering which are only sensitive to
protons. If the Au nucleus has a neutron skin, then the hadronic
radius may be larger than that quoted in~\cite{devries}. We estimated
the systematic errors by varying the Woods-Saxon parameters within the
range allowed by electron scattering data. Although unmeasured,
theoretical guidance suggests the neutron skin may add 0.2 fm to the
radius of heavy nuclei~\cite{halo}. To account for a possible neutron skin, we
increased the radius of the Au nucleus to 6.7 fm. We find that our
results only weakly depend on the radius and depend mostly on the
diffuseness parameter "a". The effect of a neutron skin is therefore
well within our quoted systematic errors and will not affect our
conclusions unless the skin significantly changes the diffuseness at
the edge of the nucleus.

 The fKLN-CGC model provides multiplicity and eccentricity. Our MCG
calculations use a two-component model and a negative binomial
distribution to estimate the event multiplicity for each simulated
event. The first parameter of the binomial distribution is generated
for each event using
\begin{linenomath}\begin{equation}
\overline{n}=f(\sqrt{s_{_{\mathrm{NN}}}})((1-x_{\mathrm{hard}}) + 2x_{\mathrm{hard}}N_{\mathrm{bin}}/N_{\mathrm{part}})
\end{equation}\end{linenomath}
where
$f(\sqrt{s_{_{\mathrm{NN}}}})=0.5933\ln(\sqrt{s_{_{\mathrm{NN}}}}/\mathrm{GeV/}c^2)-0.4153$,
$N_{\mathrm{bin}}$ is the number of nucleon-nucleon collisions,
$N_{\mathrm{part}}$ is the number of participating nucleons, and
$x_{\mathrm{hard}}$ is the fraction of the multiplicity proportional
to $N_{\mathrm{bin}}$. Then multiplicity is generated by sampling a
negative binomial distribution with parameters $\overline{n}$ and
width $k=2.1$ for each participant. This parametrization provides a
good description of multiplicity measurements in heavy-ion collisions
from $\sqrt{s_{_{\mathrm{NN}}}}=20$ to $200$ GeV~\cite{phmult} and for
all centralities. For the MCG-Q model, while the eccentricity is
defined by the locations of constituent quarks participating in the
collisions, the multiplicity is defined by the nucleon
$N_{\mathrm{part}}$ and $N_{\mathrm{bin}}$. We define the centrality
of the models according to this multiplicity so that the data and
model are treated equivalently. In this way, our eccentricity
fluctuations also contain the impact parameter and $N_{\mathrm{part}}$
fluctuations that are intrinsic to our experimental determination of a
given centrality interval. The uncertainties on the models were
estimated by varying the Woods-Saxon parameters within the range of the
errors quoted in ref.~\cite{devries}. We also varied the parameters
for the multiplicity but the results were not very sensitive to those.

\begin{figure}[htbp] \centering\mbox{
    \includegraphics[width=0.45\textwidth]{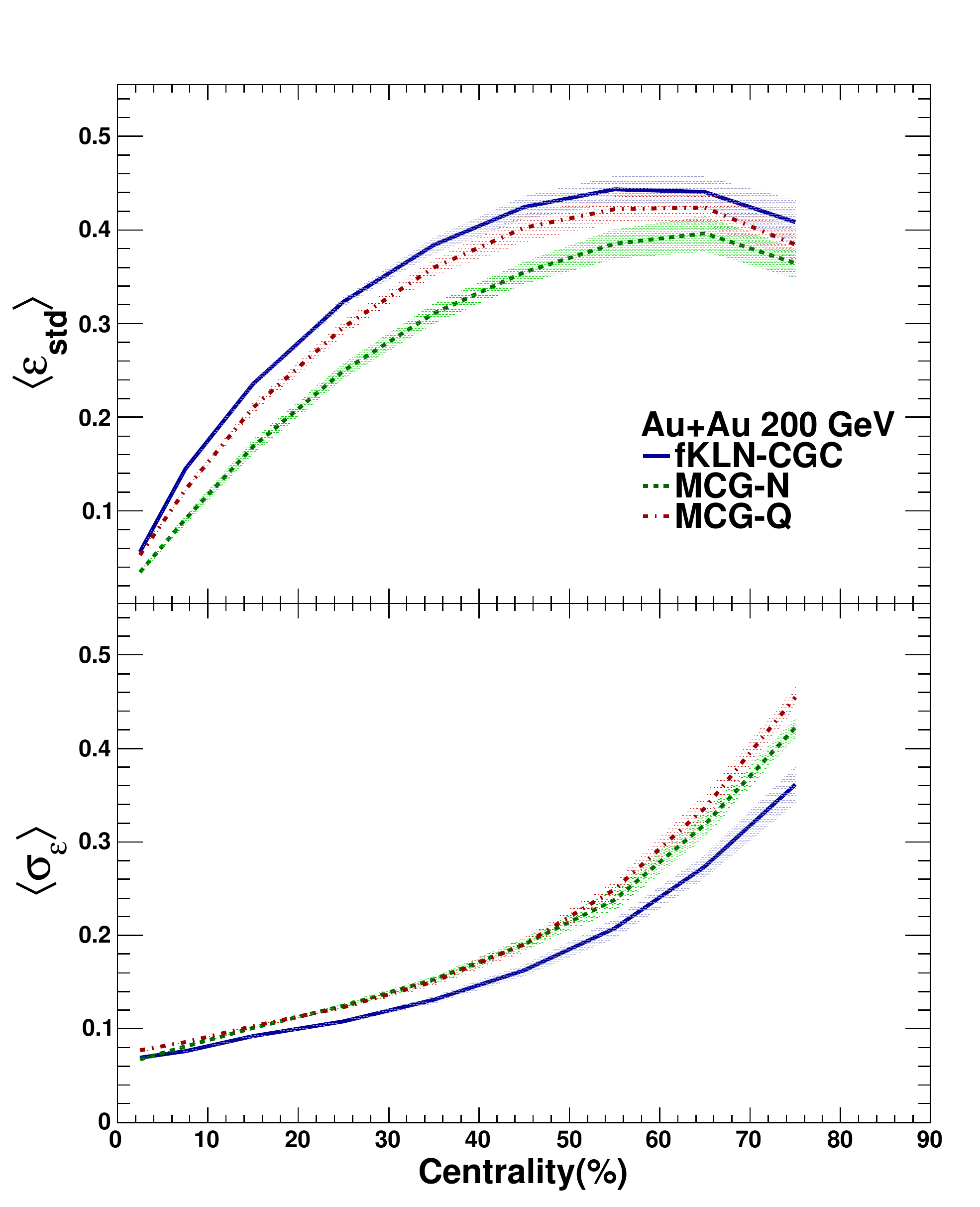}}
  \caption{$\varepsilon_{\mathrm{std}}$ (top) and $\sigma_{\mathrm{\varepsilon}}$ 
(bottom) vs. centrality for Au+Au 200
    GeV between Monte Carlo Glauber-Nucleon Participants, Monte Carlo
    Glauber-Quark Constituents, Color Glass Condensate models. The
    shaded regions show the systematic errors. } \label{fig23}
\end{figure}

Several different variables related to the eccentricity have been
calculated from the three models. This includes the eccentricity
relative to the reaction-plane ($\varepsilon_{\mathrm{std}}=\frac{\langle
  y-x\rangle}{\langle y+x\rangle}$), the eccentricity relative to the
participant plane ($\varepsilon_{\mathrm{part}}$), and the two- and four-particle
cumulants of $\varepsilon_{\mathrm{part}}$~\cite{Bhalerao:2006tp}:
\begin{eqnarray}
\varepsilon\{2\} &=& \sqrt{\langle\varepsilon_{\mathrm{part}}^2\rangle} \\
\varepsilon\{4\} &=& \left(2\langle\varepsilon_{\mathrm{part}}^2\rangle^2-\langle\varepsilon_{\mathrm{part}}^4\rangle\right)^{1/4} .
\end{eqnarray}
$\varepsilon_{\mathrm{std}}$ for 200 GeV Au+Au collisions is shown in
Fig.~\ref{fig23} (top). $\varepsilon_{\mathrm{std}}$ is largest for the fKLN-CGC
model and smallest in the MCG-N model. The MCG-Q model is intermediate
between the two. The relevant quantities have been tabulated
online~\cite{onlinetabs}. 


Figure~\ref{fig23} (bottom) shows the fluctuations of $\varepsilon_{\mathrm
  {std}}$ for the three models for 200 GeV Au+Au collisions. The
fluctuations in the two Glauber models are larger than those for the
fKLN-CGC model. One might expect the MCG-Q model to have smaller
fluctuations than the MCG-N model since there are three times as many
possible participants. This is counterbalanced however by two effects
1) the three constituent quarks are confined inside nucleons thus
inducing correlations that partially offset the effect of more
participants, and 2) the mean value of the eccentricity is larger in
the MCG-Q model. These effects lead to the result that the width of
the eccentricity distribution in the MCG-Q model and the MCG-N model
are similar. On the other hand, since the MCG-Q model gives a larger
average eccentricity, when considering
$\sigma_\varepsilon/\varepsilon$, the MCG-Q model is intermediate
between the fKLN-CGC and the MCG-N models as one might naively expect.

The trends for Cu+Cu collisions remain the same as for Au+Au
collisions with the fKLN-CGC model having the largest eccentricity and
smallest fluctuations and the MCG-Q model intermediate between the
MCG-N and fKLN-CGC models. None of the models showed a significant
difference between $\sqrt{s_{_{\mathrm{NN}}}}=62.4$ and 200 GeV, so we
only show the 200 GeV results
here. 


\begin{figure}[htbp] \centering\mbox{
   \includegraphics[width=0.45\textwidth]{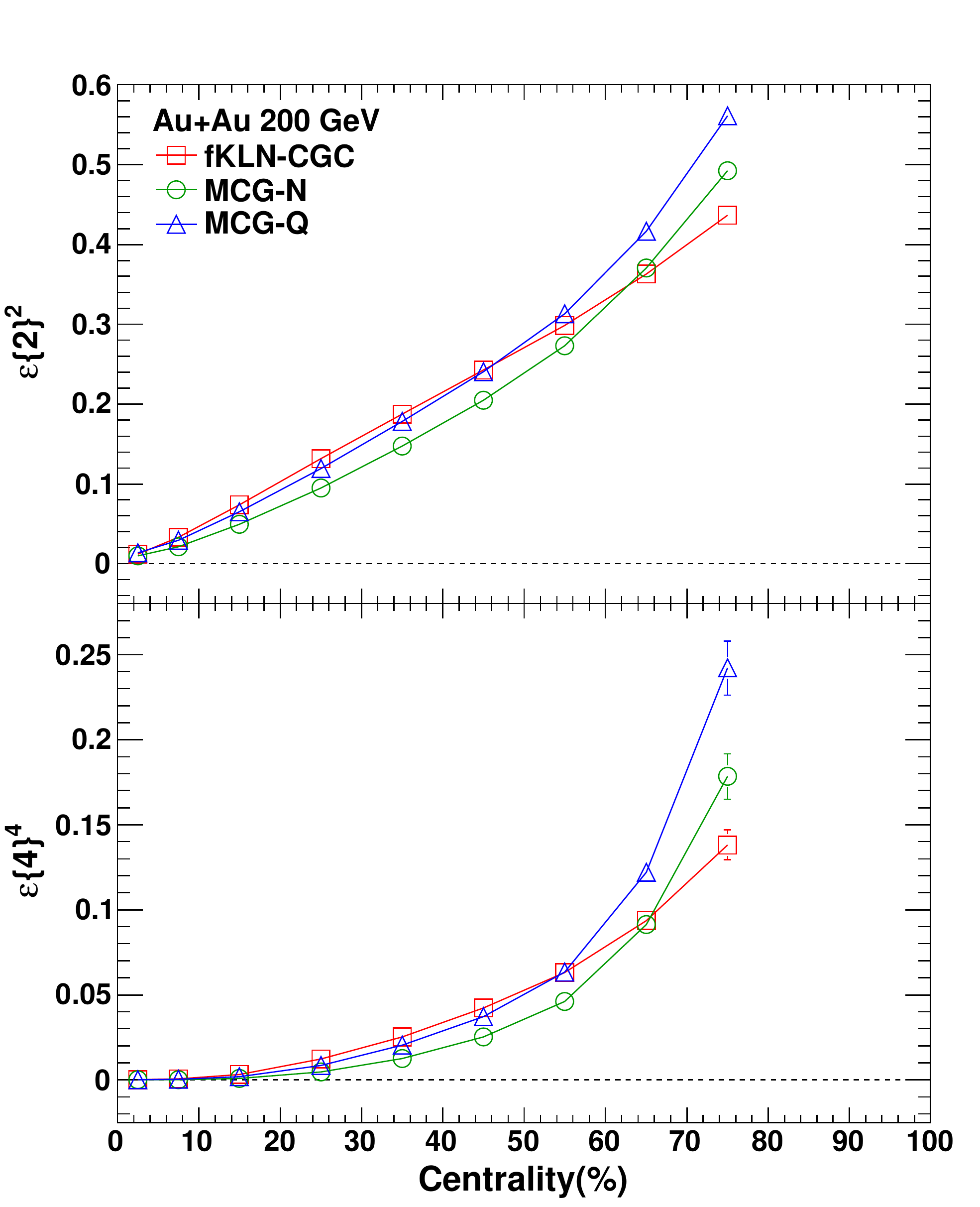}}
 \caption{$\varepsilon \{2\}^{2} $ and $\varepsilon\{4\}^{4}$ vs.
   centrality for Monte Carlo Glauber models with nucleon or
   constituent quark pariticipants and for a Color Glass Condensate
   model. }\label{fig25}
\end{figure}

 Figure~\ref{fig25} shows $\varepsilon\{2\}^{2}$ (top) and
$\varepsilon\{4\}^{4}$ (bottom) for Au+Au 200 GeV for the three
models. $\varepsilon\{2\}^{2}$ shows positive values throughout the
range and decreases with increasing centrality. The MCG-N model shows smaller
values than the other two models for central and mid-central
collisions but cross fKLN-CGC for the most peripheral collisions.
MCG-Q and fKLN-CGC models show the same values for
$\varepsilon\{2\}^{2}$ for central and mid-central collisions but
MCG-Q shows the highest values in all the three models for the most
peripheral collisions. $\varepsilon\{4\}^{4}$ shows similar behavior
as $\varepsilon\{2\}^{2}$ but it becomes negative for the most central
collisions in all the models like $v_{2}\{4\}^{4}$ for the most
central collisions in the data. This behavior is the same for Cu+Cu
collisions and different energies. In the models, this negative value
can be traced to $N_{\mathrm{part}}$ fluctuations
present when using multiplicity to select centrality bins. If
$N_{\mathrm{part}}$ is used to define the centrality in the models,
then $\varepsilon\{4\}^{4}$ remains positive, even for central
collisions.

 
\begin{table}[ht]
\caption{ The $\langle dN_{\mathrm{ch}}/d\eta\rangle$~\cite{spec}, $N_{\mathrm{part}}$ and $N_{\mathrm{bin}}$ values
  corresponding to the centrality intervals used in this paper. }
\begin{center}
\begin{tabular}{ c  c  c    c c }
  \hline
    \hline
Centrality ($\%$) & $\langle dN_{\mathrm{ch}}/d\eta\rangle$ &  $N_{\mathrm{part}}$&$N_{\mathrm{bin}}$ \\
   \hline
    \multicolumn{4}{c}{\bf Au+Au 200 GeV} \\
    \hline
    70-80 $\%$ & 22 $\pm$2 &  13.46 $\pm$0.50& 12.45 $\pm$ 0.69 \\
    60-70 $\%$ & 45 $\pm$3 & 26.62$\pm$ 0.95  &29.33 $\pm$ 1.75\\
    50-60 $\%$ & 78 $\pm$6 &  47.06 $\pm$1.21& 62.1 $\pm$ 2.1 \\
    40-50 $\%$ & 126 $\pm$9 &  75.58$\pm$1.56 & 121.8$\pm$ 4.2\\
    30-40 $\%$ & 195 $\pm$14 &  114.81$\pm$1.73 & 218.9 $\pm$ 6.1\\
    20-30 $\%$ & 287 $\pm$20 &  166.85$\pm$1.33 & 371.3 $\pm$ 6.2\\
    10-20 $\%$ & 421 $\pm$30 &  234.49$\pm$0.84 & 599.6 $\pm$ 4.5\\
    5-10 $\%$ & 558 $\pm$40 &  299.47$\pm$0.75 &845.6 $\pm$ 3.2\\
    0-5 $\%$ & 691 $\pm$49 &  349.09$\pm$0.30 & 1059 $\pm$ 3.\\
           \hline
   \multicolumn{4}{c}{\bf Au+Au 62.4 GeV} \\
   \hline
    70-80 $\%$ & 13.9 $\pm$1.1 &  13.18 $\pm$0.71 & 11.6 $\pm$ 0.87\\
    60-70 $\%$ & 29.1 $\pm$2.2 &   25.56$\pm$ 1.11 & 26.69 $\pm$ 1.87\\
    50-60 $\%$ & 53.1 $\pm$4.2 &  44.97 $\pm$1.27 &55.9$\pm$ 2.9\\
    40-50 $\%$ & 87.2 $\pm$7.1 &  72.70$\pm$1.25 & 107.4 $\pm$ 3.8\\
    30-40 $\%$ & 135 $\pm$11 & 110.53$\pm$1.05 & 190.5 $\pm$ 4.5\\
    20-30 $\%$ & 202 $\pm$17 &  161.08$\pm$0.97 & 319.4 $\pm$ 4.8\\
    10-20 $\%$ & 292 $\pm$25 &  228.51$\pm$0.52 & 514.6 $\pm$ 3.4\\
    5-10 $\%$ & 385 $\pm$33 & 293.39$\pm$0.96 & 721.7 $\pm$ 3.9\\
    0-5 $\%$ & 472 $\pm$41 &   343.82$\pm$0.44& 900.75 $\pm$ 1.85\\
    \hline
       \multicolumn{4}{c}{\bf Cu+Cu 200 GeV} \\
   \hline
    50-60 $\%$ & 25.3 $\pm$1.6 &  16.41$\pm$0.24 &15.71$\pm$0.31 \\
    40-50 $\%$ & 38.7 $\pm$2.5 &  25.14$\pm$0.16& 27.42 $\pm$ 0.22 \\
    30-40 $\%$ & 56.9 $\pm$3.7 &  37.35$\pm$0.47 & 46.87 $\pm$ 1.00\\
    20-30 $\%$ & 82.9 $\pm$5.4 &  53.07$\pm$0.29 &75.46 $\pm$ 0.42\\
    10-20 $\%$ & 119 $\pm$7.7 &  73.61$\pm$0.12 &119.65 $\pm$0.15\\
    0-10 $\%$ & 170 $\pm$11 &  98.08$\pm$0.11& 182.7 $\pm$ 0.30 \\
    \hline
           \multicolumn{4}{c}{\bf Cu+Cu 62.4 GeV} \\
   \hline
    50-60 $\%$ & 17.4 $\pm$1.1 &  15.36$\pm$0.05& 13.91$\pm$0.01 \\
    40-50 $\%$ & 26.3 $\pm$1.7 &  23.92$\pm$0.05 & 25.56$\pm$0.06\\
    30-40 $\%$ & 38.7 $\pm$2.5 &  35.62$\pm$0.05 &41.09$\pm$ 0.12\\
    20-30 $\%$ & 56.4 $\pm$3.7 &  50.76$\pm$0.12 &65.86$\pm$0.30\\
    10-20 $\%$ & 81.2 $\pm$5.3 &  70.67$\pm$0.50 &103.15$\pm$0.95\\
    0-10 $\%$ & 117 $\pm$7.7 &  94.98$\pm$0.25&155.65$\pm$0.75 \\
    \hline
        \hline
          \end{tabular}
\end{center}
\label{cent}
\end{table}%

\end{document}